\documentclass[12pt,aps,showpacs,preprintnumbers,onecolumn,superscriptaddress,floatfix,nofootinbib]{article}
\usepackage{mathtools}
\usepackage{authblk}
\usepackage{amsmath,amsfonts,amssymb}
\usepackage{graphicx} 
\usepackage{algorithm}
\usepackage{algorithmicx}
\usepackage{algpseudocode}
\usepackage{xcolor}
\usepackage{comment}
\usepackage{appendix}
\usepackage{bm}

\title{An exact and fast solution of the inverse Regularized Optimal Transport problem}

\author[1*]{Dario Mazzilli}
\author[1]{Riccardo Piombo}
\author[1]{Lorenzo Buffa}
\author[1]{Aurelio Patelli}
\affil[1]{Enrico Fermi Research Center, 00184 Rome}
\affil[*]{dario.mazzilli@cref.it}
\usepackage[sorting=none,style=numeric,backend=biber, giveninits=true, maxbibnames=1, minnames=1]{biblatex} 
\begin{document}

\maketitle

\begin{abstract}
 Optimal transport describes the most efficient way to move mass between two distributions, given a cost matrix for moving mass between each pair of locations. Entropic optimal transport, solved via the Sinkhorn algorithm, is a widely used regularized version of this problem. Its inverse problem asks the opposite question: given an observed transport plan, what cost matrix produced it? This is difficult because the cost is identifiable only up to an additive gauge freedom. Here we show that this freedom can be fixed exactly by a single double-centering operation applied to the observed plan, yielding the true cost matrix in closed form, with no iterative optimization required. When a modest number of true cost entries are known, the same approach lets us jointly estimate the temperature parameter controlling the entropic regularization, together with a diagnostic for the reliability of this estimate. We further show that the method is not specific to the entropic optimal transport, but extends to a broader class of transport models defined by an invertible relation between cost and plan.
\end{abstract}

\section*{Introduction}
\label{sec:intro}

The theory of \emph{Optimal Transport} (OT) provides a fundamental 
mathematical framework for comparing probability distributions through 
a cost-minimizing coupling.
Originally formulated by Monge \cite{monge1781memoire} and rigorously relaxed by 
Kantorovich \cite{kantorovich1942translocation}, OT seeks the most economical way to transfer one 
distribution of mass onto another, given a cost matrix whose entries 
quantify the expense of moving a unit of mass from each source to each 
target.
The OT problem has been framed in many different systems, spanning 
developmental biology and single-cell genomics~\cite{schiebinger2019optimal}, 
ecology~\cite{stock2021optimal}, economics~\cite{galichon2016optimal}, 
transportation networks~\cite{ibrahim2021optimal,bernot2008optimal}, computer 
vision~\cite{rubner2000earth}, and natural language 
processing~\cite{kusner2015word}.

While the \emph{forward} OT problem seeks the optimal transport plan 
for a given cost, the \emph{inverse} OT problem asks the opposite 
question: given an observed optimal transport plan, what cost matrices 
could have produced it as an OT minimizer?
This inverse perspective has been extensively studied and appears naturally in diverse domains, such as 
economics, transportation networks, 
 machine learning and ecology \cite{galichon2016optimal,bernot2008optimal,ibrahim2021optimal,peyre2019computational,Young2021,cuturi2014ground,kerdoncuff2021metric,Li2018,stuart2020inverse,Chiu2022,Dupuy2019}.

The inverse problem is fundamentally underdetermined, because an entire 
class of cost matrices can equally explain the observed plan. 
This \emph{gauge freedom} means that a cost matrix is identifiable only 
up to an equivalence class, and any meaningful comparison or recovery 
of costs must account for it explicitly.

A prominent and practically convenient variant of OT is 
\emph{Entropic Optimal Transport} (EOT), which regularizes the 
classical problem by adding a weighted Shannon entropy term to the 
objective~\cite{Cuturi2013}. 
This regularization, controlled by a temperature parameter 
\(\varepsilon\), makes the problem strictly convex, and gives rise to efficient computational schemes based on 
iterative scaling, most notably the Sinkhorn 
algorithm~\cite{Sinkhorn1967}.
The optimal entropic plan inherits a gauge structure closely analogous 
to that of classical OT: the transport plan is invariant under 
transformations of the cost matrix that add arbitrary source- and 
target-specific terms, exactly as in the unregularized case, except 
that in EOT the relevant equivalence class is defined up to additive row and 
column terms \emph{rescaled by the temperature} \(\varepsilon\) (following standard literature, we refer to this parameter as a temperature, although it lacks a thermodynamic interpretation).
In the zero-temperature limit \(\varepsilon \to 0\), the entropic 
plan concentrates on the support of a classical OT solution and the 
two gauge structures coincide.

A natural and principled way to fix the gauge is \emph{double centering}, a transformation that selects a canonical representative of each equivalence class.
It therefore isolates the purely interaction component of the cost --- the part that governs pairwise "affinity" between sources and targets, 
net of any individual source or target bias.
We show that the original cost matrix can be recovered exactly, up to gauge equivalence, 
and that double centering yields a unique and geometrically canonical reconstruction. 

Recent works have addressed the inverse OT problem through several 
complementary viewpoints. Polyhedral and combinatorial 
approaches~\cite{liu2013perturbation} study the structure of normal 
cones and feasible perturbations of the cost.
Differentiable OT and Sinkhorn-based methods~\cite{genevay2018learning,
Chiu2022} exploit the smooth dependence of the entropic OT objective 
on the cost, enabling efficient gradient-based cost learning.
Several other strategies assume a functional form for the 
cost~\cite{Dupuy2019}, or propose variational formulations based on 
entropic regularization~\cite{Ma2020,Li2018} and probabilistic 
approaches that characterize the set of admissible costs compatible 
with empirical couplings~\cite{Chiu2022,Carlier2022}. A complementary 
line of work treats cost inference as a Bayesian, stochastic-
optimization problem, recovering costs from noisy observations of the 
optimal plan using only a forward OT solver~\cite{stuart2020inverse}; 
more recently, well-posedness and stability results originally 
established for entropic regularization~\cite{Chiu2022} have been 
extended to general Bregman regularizers, broadening the class of 
link functions for which the inverse problem is provably 
well-posed~\cite{bao2026well}.

Unlike these approaches, which typically recover $C$ through an 
iterative optimization loop -- gradient descent through a 
differentiable OT solver, Bayesian sampling with a forward solver at 
each step, or alternating minimization -- the method we propose here 
yields an exact, closed-form solution: a single double-centering 
operation applied to a simple transform of the observed plan, with no 
optimization loop and no repeated calls to a forward OT solver. 
This has two direct consequences. 

First, our recovery makes no assumption on the structure of $C$: unlike methods that impose metric, low-rank, sparsity, or parametric constraints on the cost to obtain a unique estimate, our approach recovers a completely generic cost matrix exactly, up to the intrinsic gauge freedom of the forward problem. This structure-agnostic character is particularly relevant when the effective cost cannot be reliably assigned to a prescribed class a priori. In many real systems, costs may depend simultaneously on geographical, economic, institutional, and network-specific factors, so that physical distance represents only one component of the effective interaction cost. Airline fares provide a simple example, as their structure depends not only on route distance but also on hub organization, competition, and market power~\cite{borenstein1989hubs}. Avoiding structural priors on $C$ therefore allows the inferred pairwise cost structure to be determined by the observed transport plan rather than forced into a preselected functional family. Our approach further allows the temperature $\varepsilon$ to be estimated, whenever a modest number of true cost entries are available.

Second, the closed-form 
nature of the estimator makes it computationally inexpensive: 
recovering $C$ from an observed plan requires only a handful of matrix 
operations, in contrast to the many forward-solver evaluations 
required by iterative schemes. This computational advantage is 
expected to become increasingly relevant as the size of the 
transportation problem grows, since a single row/column centering 
scales far more favorably than repeated Sinkhorn iterations or 
gradient-based optimization over the cost matrix.

\section*{Methods}
\subsection*{The Entropic Optimal Transport}
Formally, in the discrete case, the Kantorovich problem reads:
\begin{equation}
\label{eq:classical_ot}
\mathcal{T}_C(s, r)
= \min_{W \in \Pi(s, r)} \langle C, W \rangle
= \min_{W \ge 0}
\sum_{i,j} C_{ij} W_{ij}
\quad
\text{s.t. } W \mathbf{1} = s, \;
W^\top \mathbf{1} = r,
\end{equation}
where \( \Pi(s, r) \) denotes the set of feasible couplings with marginals \( s \) and \( r \).  
The minimal cost \( \mathcal{T}_C(s, r) \) defines the \emph{optimal transport cost}.

The dual formulation of OT reveals its convex-analytic and geometric structure.  
Introducing potentials \( f \) and \( g \), one obtains:
\begin{equation}
\label{eq:dual_ot}
\mathcal{T}_C(s, r)
= \max_{f, g}
\Big\{
\langle f, s \rangle + \langle g, r \rangle
:\;
f_i + g_j \le C_{ij}, \; \forall i,j
\Big\}.
\end{equation}  
In geometric terms, the feasible set of couplings forms a convex polytope, and each optimal plan corresponds to a vertex where the linear functional defined by \( C \) attains its minimum.

Despite its elegant structure, the classical OT problem is computationally demanding: the optimization is high-dimensional and the feasible region is constrained and non-smooth.  
To address these issues, \emph{entropic regularization}~\cite{Cuturi2013} modifies the objective by adding a strictly convex term:
\begin{equation}
\label{eq:entropic_ot}
\mathcal{T}_{C,\varepsilon}(s, r)
= \min_{W \in \Pi(s, r)}
\langle C, W \rangle
+ \varepsilon \, \mathrm{KL}(W \,\|\, s \otimes r),
\end{equation}
where \( \varepsilon > 0 \) controls the regularization strength and
\[
\mathrm{KL}(W \,\|\, s \otimes r)
= \sum_{i,j} W_{ij}
\log\frac{W_{ij}}{s_i r_j}
\]
is the Kullback--Leibler divergence.  

The entropic term ensures strict convexity, yielding a unique optimal plan \( W^\varepsilon \) with full support.  
This regularized formulation admits an efficient solution via \emph{Sinkhorn's algorithm}~\cite{Sinkhorn1967,Cuturi2013}, an iterative matrix-scaling procedure exploiting the separable structure of the constraints.

The dual problem becomes smooth:
\begin{equation}
\label{eq:dual_entropic_ot}
\mathcal{T}_{C,\varepsilon}(s, r)
= \max_{f, g}
\Big\{
\langle f, s \rangle + \langle g, r \rangle
- \varepsilon \sum_{i,j} s_i r_j
\exp\!\Big(\frac{f_i + g_j - C_{ij}}{\varepsilon}\Big)
\Big\},
\end{equation}
and its optimality conditions give the Gibbs form of the optimal coupling:
\begin{equation}
\label{eq:sinkhorn}
W_{ij}^\varepsilon
= s_i r_j
\exp\!\Big(\frac{f_i + g_j - C_{ij}}{\varepsilon}\Big).
\end{equation}
As \( \varepsilon \to 0 \), the regularized solution \( W^\varepsilon \) converges to a sparse optimal plan \( W^\star \) of the standard OT problem, while for finite \( \varepsilon \), the regularization smooths the polyhedral geometry into a differentiable manifold structure.

\subsection*{The Inverse Optimal Transport Problem}
In OT and EOT formulations, the cost matrix plays a central role: together with the marginal constraints, it defines the primary input of the optimization problem. This setting is appropriate when costs are directly measurable, as in systems where physical distances or energy expenditures quantify transport. 

In many real-world networks, however, the situation may be reversed. In ecological or trade systems, for example, optimization is presumed to occur, yet the underlying cost function is not directly observable, what can be measured instead is the resulting transport plan \cite{Young2021}. 

In such contexts one  may want to estimate the 'coupling' matric $C$ that ruled the optimization.
However, symmetries deeply influence the forward problems and, as a consequence, constrain their corresponding inverse formulations. In classical OT, several transformations of the cost matrix leave the optimal transport plan invariant such as global affine transformations of the form 
\begin{equation}
\label{eq:affine_transf}
\bar{C} = \lambda\,C + \bm{a} \otimes \mathbf{1} + \mathbf{1} \otimes \bm{b} + \gamma\,\mathbf{1} \otimes \mathbf{1}.
\end{equation}
The operator $\otimes$ is the outer product of two vectors, $\lambda > 0$ is a scale constant, $\bm{a} \in \mathbb{R}^n$ and $\bm{b} \in \mathbb{R}^m$ are two arbitrary vectors, and $\gamma \in \mathbb{R}$ sets the "zero" of the costs. Element-wise, Eq.~(\ref{eq:affine_transf}) reads as
\begin{equation}
\bar{C}_{i\alpha} = \lambda C_{i \alpha} + a_i + b_\alpha + \gamma
\end{equation}
Such transformations can be fully absorbed by redefining the corresponding dual potentials $u_i$ and $v_\alpha$ as
\begin{equation}
    \bar{u}_i = u_i - a_i \qquad \bar{v}_\alpha = v_\alpha - b_\alpha
\end{equation}
In this way, the dual constraints $u_i + v_\alpha \le C_{i\alpha}$ and the equality conditions on the support of the optimal plan remains unchanged. Consequently, both the dual geometry and the entire set of optimal transport plans remain invariant under the transformation in Eq.~\eqref{eq:affine_transf}.

OT also exhibits \emph{local} symmetries that affect specific optimal solutions. These solutions correspond to vertices of the transport polytope and are associated with a normal cone that defines a family of cost perturbations leaving the same transport plan optimal:  perturbing the cost within this cone is analogous to a limited rotation in the space of admissible cost directions \cite{Villani2009}.

In the EOT framework the presence of entropy removes the discrete structure of normal cones and replaces local invariance with a smooth, differentiable dependence of the transport plan on the cost matrix: small variations in the cost lead to smooth adjustments of the transport plan, ensuring continuous sensitivity to the underlying cost structure \cite{Villani2009}. However, the invariance in Eq.~(\ref{eq:affine_transf}) persist except for the rescaling of costs ($\lambda$ term), which would alter the strength of the entropic regularization.

To illustrate this, consider the global additive transformation of the cost matrix in Eq.~(\ref{eq:affine_transf}), except for the constant term $\gamma$, which can be fixed afterwards being a uniform offset that does not affect the solution. This transformation can be compactly written as
\begin{equation}
\label{eq:invariance_subot}
\bar{C} = C + \bm{a} \oplus \bm{b}
\end{equation}
where the operator $\oplus$ denotes the outer sum $(\bm{a} \oplus \bm{b})_{i\alpha} = a_i + b_\alpha$.  In EOT, this invariance appears at the level of the Gibbs kernel \(K_{ij} = e^{-C_{ij}}\) as a row--column diagonal scaling:
\begin{equation}
\label{eq:KbarDrKDc}
\bar K_{ij} := e^{-\bar C_{ij}}
= e^{-(C_{ij}+a_i+b_j)}
= e^{-a_i}\, K_{ij}\, e^{-b_j}
= (D_r\, K\, D_c)_{ij}
\end{equation}
with \(D_r = \mathrm{diag}\,(a_1, \dots, a_n)\) and \(D_c = \mathrm{diag}\,(b_1,\dots,b_m)\). 

\section*{Results}
\subsection*{Fixing the gauge}
The invariance in Eq.(\ref{eq:invariance_subot}) implies that a single transport plan can be linked to multiple cost matrices. The set of all such matrices defines an equivalence class
\begin{equation}
\label{eq:equiv_class}
[\![C]\!] \sim \{\, \bar{C} \mid \bm{a} \in \mathbb{R}^n, \bm{b} \in \mathbb{R}^m \,\}
\end{equation}
so that any two elements of this set differ only by a separable component depending on row and column features. 

Within this class, the absolute reference of the cost function is undefined, which raises the question of how to select a representative cost matrix. To resolve this ambiguity, we exploit the \emph{double-centering operation}, which removes the additive row and column contributions from any cost matrix and projects it onto a unique representative of its equivalence class. The double-centered matrix $\bar{C}'$ is defined as
\begin{equation}
\label{eq:dc}
 \bar{C}'_{i\alpha}
= \bar{C}_{i\alpha}
- \frac{1}{N} \sum_{k} \bar{C}_{k\alpha}
- \frac{1}{M} \sum_{l} \bar{C}_{i l}
+ \frac{1}{MN} \sum_{k l} \bar{C}_{k l}   
\end{equation}
This operation subtracts the average of each column and row and then restores the global mean, ensuring that both row and column averages of $\bar{C}'$ vanish. Performing the algebra shows that 
\begin{equation}
\bar{C}'_{i\alpha} = C'_{i\alpha} \qquad \forall\, \bm{a} \in \mathbb{R}^n \quad \forall\, \bm{b} \in \mathbb{R}^m
\end{equation}
Consequently, the double-centering operation uniquely identifies the equivalence class $[\![\bar{C}]\!]$ by fixing the arbitrariness associated with the additive gauge that characterize the class in Eq.(\ref{eq:invariance_subot}). In this way, each transport plan is associated with a unique and well-defined cost matrix within its equivalence class.

\subsection*{An exact solution of inverse problem}
The problem requires to infer, from an observed transport plan $W$, the cost matrix $C$ that was used to obtain such plan. In principle, one could solve this problem with linear programming, decomposing the plan $W$ in a 'true' coupling contribution plus two components of rank 1. In practice, the solution is much more straightforward.
Having chosen the double centered matrix $C'$ as the matrix representative of the entire class associated to an observed transportation plan $W$, the inverse problem is immediately solved. For example, in EOT, the matrix $C^\prime$ is simply obtained
\begin{equation}
\label{eq:estimator}
    C' = (-\log W)'
\end{equation}
that follows from eq.\ref{eq:sinkhorn} and the definition of double centering eq.\ref{eq:dc}.

More generally, the gauge symmetry underlying Eq.~\eqref{eq:estimator} is not a specific property of entropic optimal transport. 
It follows directly from the structure of any transport problem in which the cost enters linearly and the row and column marginals of the plan are constrained. 
Indeed, consider an objective of the generic form
\begin{equation}
\mathcal{L}(W)
=
\sum_{ij} C_{ij}W_{ij}
+ \mathcal{R}(W),
\end{equation}
where $\mathcal{R}(W)$ contains any additional regularization or entropic contribution, while the marginals of $W$ are fixed. 
Under the transformation in \eqref{eq:invariance_subot} the objective changes, for every feasible plan, only by $\sum_i a_i s_i+\sum_j b_j r_j$, which is independent of $W$. 
Therefore, the minimizing transport plan is unchanged. 
The additive row-column gauge is thus a generic consequence of the linear cost term and the marginal constraints, rather than a peculiarity of the Sinkhorn formulation.
This observation applies in particular to several regularized and statistical-mechanical formulations of transport, including EOT, the SubOT model~\cite{buffa2025maximum}, the ensemble formulation of Ref.~\cite{baybusinov2026grand}, and other regularized extensions such as Ref.~\cite{muzellec2017tsallis}. 
In all these cases, the row and column constraints are enforced through Lagrange multipliers, which inherit precisely the same additive structure as the gauge transformations of the cost.

For a broad subclass of such models, this symmetry can also be exploited to obtain an explicit solution of the inverse problem. 
If the stationarity condition with respect to each entry of the plan can be written in the form
\begin{equation}
    W_{ij} = F\left(C_{ij}+f_i+g_j\right),
    \label{eq:general_link}
\end{equation}
where $F$ is a model-dependent invertible link function and $f_i$ and $g_j$ are the Lagrange multipliers associated with the marginal constraints, then
\begin{equation}
    F^{-1}(W_{ij}) = C_{ij}+f_i+g_j.
\end{equation}
Double centering therefore eliminates the dual contributions and gives
\begin{equation}
    \bigl(F^{-1}(W)\bigr)'=C',
\end{equation}
up to any model-dependent overall scale. For EOT the inverse link is logarithmic, while for SubOT it is reciprocal. 
The same construction applies whenever the corresponding stationarity relation admits such an invertible entry-wise link.

In this paper, if not elsewhere specified, we perform our simulations and analysis for the EOT model, introduced by Cuturi et al. \cite{Cuturi2013}.

\paragraph{Robustness to noise}

Real-world transport plans are rarely observed exactly. Empirical 
flow data --- trade volumes, migration counts, mobility traces, or 
ecological interaction networks --- are subject to measurement error, 
finite-sample fluctuations, and reporting biases that corrupt the 
observed plan $W$ relative to the true underlying coupling. A particularly relevant class of measurement errors are 
\emph{node-wise} multiplicative perturbations, in which each entry 
of the observed plan is corrupted by source- and target-specific 
factors:
\begin{equation}
    \widetilde{W}_{ij} = W_{ij} \cdot \alpha_i \cdot \beta_j,
    \label{eq:nodewise}
\end{equation}
for arbitrary strictly positive vectors $(\alpha_i)$ and $(\beta_j)$.
Such errors arise naturally whenever the marginal distributions $s$ 
and $\sigma$ are themselves measured with multiplicative noise, or 
whenever the observation process introduces systematic row- and 
column-specific biases --- for instance, heterogeneous detection 
efficiencies across sources and targets.

In log-space, perturbation~\eqref{eq:nodewise} takes the additive 
form $\log \widetilde{W}_{ij} = \log W_{ij} + \log \alpha_i + 
\log \beta_j$, which lies entirely in the gauge subspace spanned by 
source- and target-specific additive terms (see Appendix A).
Since double centering projects orthogonally onto the complement of 
this subspace, it annihilates the perturbation exactly, and the 
estimator satisfies
\begin{equation}
    \widetilde{C}' = -(\log \widetilde{W})' = -(\log W)' = C'
\end{equation}
identically, regardless of the magnitude of $\alpha_i$ and $\beta_j$.
The recovery of $C'$ is therefore \emph{exactly} robust to node-wise 
multiplicative errors, not merely approximately so: even if the 
marginals $s$ and $\sigma$ are arbitrarily misspecified, the 
interaction structure of the cost matrix remains perfectly 
recoverable from the observed plan.

This invariance is not merely a technical property of the estimator, but is closely related to the structure of the transport problem itself. 
In both optimal transport, source- and target-specific properties are already encoded through the marginal constraints and their associated dual parameters. 
The cost matrix is therefore identifiable only through its non-separable component, namely the part that describes how specific source-target pairs deviate from what could be explained by the individual properties of the two nodes alone.
The exact invariance is thus partly intrinsic to the formulation of transport models, including their ensemble counterparts: inference is driven by the joint organization of flows across pairs of nodes, rather than by variations in the overall activity, scale, or observability of individual nodes. In Appendix A we derive a full description of propagation of noise with $rank >2$.

\paragraph{Double centering does not require, nor preserve, a metric cost}

A common modeling assumption in the OT literature is that the ground 
cost $C_{ij}$ represents a genuine distance --- most typically the 
Euclidean distance, or its square, between source and target features 
--- motivated by the interpretation of the optimal transport cost as a 
Wasserstein distance between the underlying measures \cite{peyre2019computational}. We emphasize that 
this assumption plays no role in the recovery method developed here: 
double centering, and the exact identifiability result hold for an arbitrary cost matrix $C$, with no 
requirement of symmetry, non-negativity, zero diagonal, or the triangle 
inequality. The gauge freedom is a structural 
property of the OT dual problem itself, entirely independent of any 
metric interpretation of $C$.

\begin{figure}
    \centering
    \includegraphics[width=1\linewidth]{
    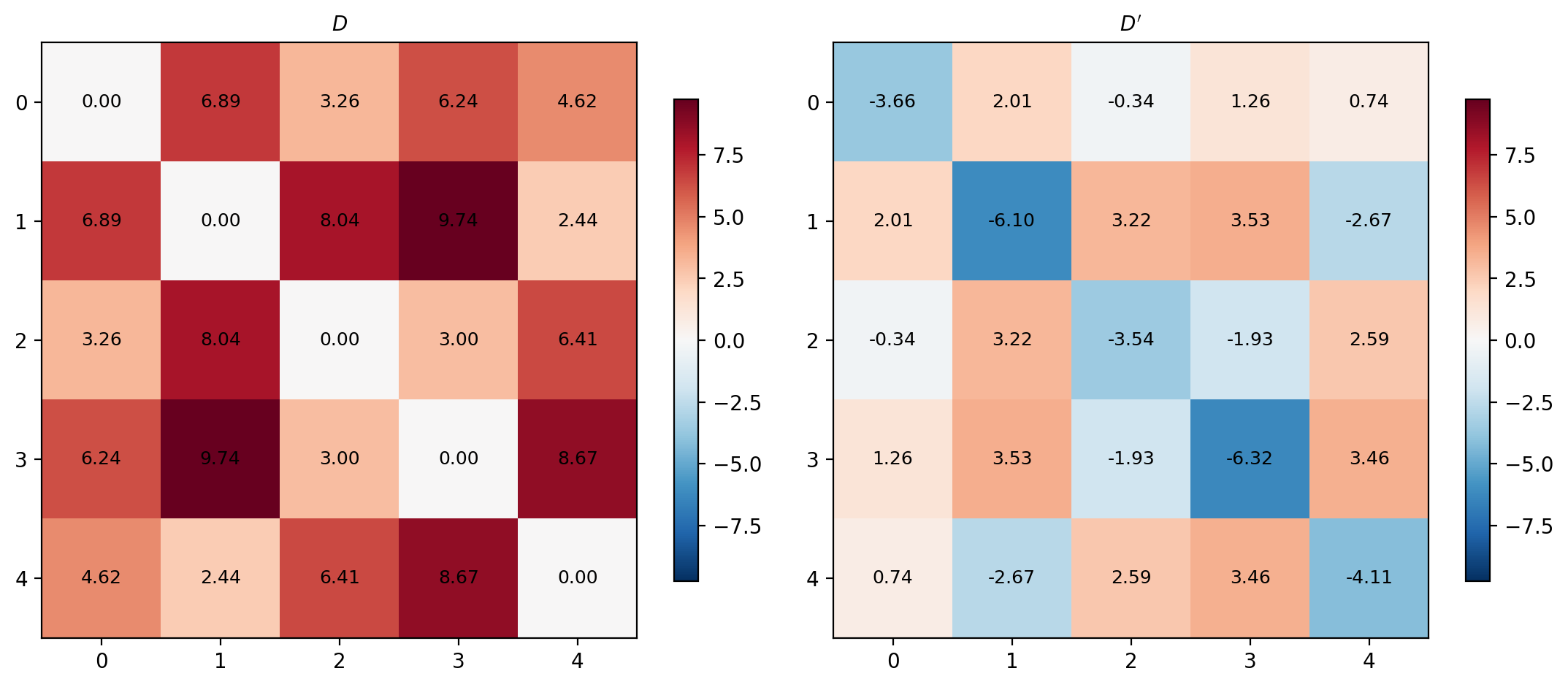}
    \caption{\textbf{Double-centering a metric cost} An Euclidean distance matrix $D$ and its double centered $D'$. While the symmetry of the metric matrix is preserved, the positivity and triangular inequality are lost.}
    \label{fig:distance_dc}
\end{figure}

In fact, the canonical, gauge-fixed representative $C'$ selected by 
double centering is generically \emph{not} a metric, even when the 
underlying true cost $C$ is. This is illustrated in 
Figure~\ref{fig:distance_dc} with a genuine Euclidean distance matrix 
$D$ computed between five points in the plane: $D$ is symmetric, 
non-negative, has zero diagonal, and satisfies the triangle inequality 
by construction (zero violations across all $n^3$ ordered triples). Its 
double-centered representative $D'$ retains the symmetry of $D$ ---  a 
general consequence of double centering applied to any symmetric input 
--- but loses every other metric property: the diagonal becomes 
non-zero and, in fact, strictly negative for every point in this 
example; roughly two-thirds of the ordered triples now violate the 
triangle inequality; and several entries are negative, which is 
incompatible with any notion of distance. 

This is not a shortcoming of the double-centering procedure, but a 
direct consequence of what double centering is designed to do: it 
isolates the component of $C$ that is invariant under the OT dual 
symmetry, discarding exactly the row- and column-wise additive terms 
that are irrelevant to the transport plan. Whatever metric structure 
$C$ may have possessed is, by construction, only partially encoded in 
this invariant component, and there is no reason to expect the 
residual to itself satisfy metric axioms. Conversely, this shows that 
imposing a metric structure on the recovered cost --- as is
done for example in the inverse OT literature to obtain a (not) unique or more 
interpretable solution~\cite{Li2018} --- is an additional, external 
modeling choice rather than a property intrinsic to the OT problem or 
required for exact gauge-fixed recovery.
Indeed, real life costs are typically not only metric. Our method makes no such 
assumption, and remains exact regardless of whether the true underlying 
cost happens to be a distance or an arbitrary, unconstrained matrix of 
pairwise affinities.

\subsubsection*{True cost recovery under partial information}

When a subset $\Omega$ of $L$ entries of the true cost matrix $C$ are 
known, one can exploit the gauge freedom to improve the reconstruction 
beyond the baseline estimate $C'$.
Specifically, since $C'$ and $C$ belong to the same gauge equivalence 
class up to the estimation error introduced by noise on $W$, one can 
search for the element $C^* = C' + f_i + g_j$ of the gauge class of 
$C'$ that best matches the known entries, by solving the linear system
\begin{equation}
    f_i + g_j = C_{ij} - C'_{ij} \quad \forall (i,j) \in \Omega
    \label{eq:partial_system}
\end{equation}
for the gauge parameters $f \in \mathbb{R}^n$ and $g \in \mathbb{R}^m$. 
This system has $n+m-1$ effective degrees of freedom, so that for $L \leq n+m-1$ observations it 
is underdetermined and admits exact solutions, while for $L > n+m-1$ 
it becomes overdetermined and is solved in the least-squares sense (see Appendix D). However, if there is no noise in the measure of $W$ or $C$, the system \ref{eq:partial_system} is always solved perfectly.

\begin{figure}[htb]
    \centering
    \includegraphics[width=0.8\linewidth]{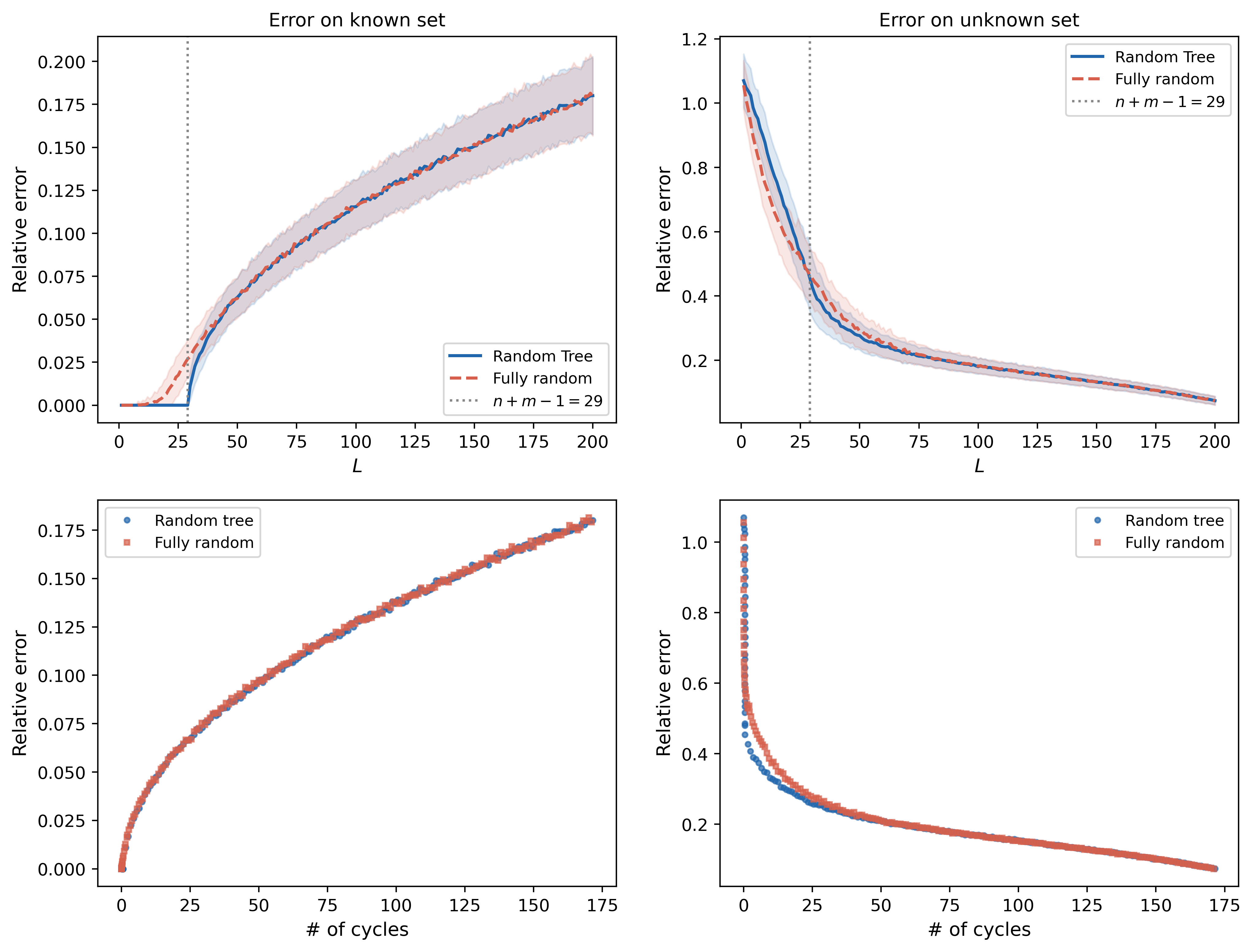}
    \caption{\textbf{Fitting prior knowledge of the cost function.} Relative error for the reconstruction of the true cost matrix, as a function of the number of known values L (upper panels) and number of cycles formed by such values (lower panelzùs). Left panels show the error on the L known values, right panels show the error on the rest of the unkown values of C. Increasing L decrease the ability to perfectly fit the known values but increase the overall precision on the unknown costs. See Appendix D for details on the setup.}
    \label{fig:partial_info}
\end{figure}

The ability to solve~\eqref{eq:partial_system}  crucially depends on 
the structure of the observed bipartite graph $G_\Omega$ induced by $\Omega$, 
whose nodes are the source and target indices and whose edges are the 
observed pairs $(i,j)$.
The condition $L \leq n+m-1$ is not sufficient to have a relative errors equal zero. If those L known entries do not span over a subset of the systems of eq. \ref{eq:partial_system} of dimensions L, meaning all independent equations, we still end up with an overdetermined system.
In terms of network topology, $G_\Omega$ need to have no cycle in order to have L independent equations. The number cycles quantify how much overdetermined the systems is.
Conversely, as long as $G_\Omega$ is a tree or a forest(no cycles), 
system~\eqref{eq:partial_system} is always exactly solvable regardless 
of $L$, and the known entries are recovered with machine precision.

Figure~\ref{fig:partial_info} illustrates these two competing effects across 
two sampling strategies: a \emph{random spanning tree} sampler, which 
guarantees zero cycles for $L \leq n+m-1$ and introduces exactly 
$L - (n+m-1)$ independent cycles beyond the threshold, and a 
\emph{fully random} sampler, which introduces cycles already for small 
$L$ (see Methods for details).
The top-left panel shows the reconstruction error on the known entries: 
the spanning-tree sampler achieves machine-precision recovery for all 
$L \leq n+m-1$, with error growing only beyond the threshold as cycles 
become unavoidable, while the fully random sampler accumulates error 
from the first cycles introduced at small $L$.
The top-right panel shows the complementary quantity: the error on the 
\emph{unknown} entries decreases monotonically with $L$ for both 
samplers, as additional known entries better constrain the gauge 
parameters $f$ and $g$ and improve the prediction of unobserved costs.
The bottom panels confirm that the number of independent cycles in 
$G_\Omega$, rather than $L$ itself, is the primary predictor of the 
fitting error on known entries, with the two sampling strategies 
collapsing onto a single curve when plotted against cycle count.
Taken together, these results show that the optimal use of partial 
cost information is achieved by sampling entries that form a spanning 
tree of the bipartite graph: this minimizes fitting error on known 
entries while maximally constraining the gauge, and the threshold 
$L = n+m-1$ marks the exact point beyond which exact recovery of the 
known entries is no longer achievable.
It should be well stressed that these results are determined under a specific form of noise in the matrix $W$ (see appendix D). Other functional form of noise may affect them. 

In appendix B we study the accuracy of the reconstruction in the complementary setting of missing values in the observed transportation plan $W$

\subsubsection*{Estimating the temperature $\varepsilon$}

The estimator derived so far recovers the gauge-fixed cost $C'$ only 
up to the unknown scale factor $\varepsilon$, since 
$C' = C/\varepsilon$ within the gauge class. 
As discussed above, $\varepsilon$ is not identifiable from the 
transport plan alone: any rescaling of $\varepsilon$ can be absorbed 
into a corresponding rescaling of $C$, leaving $W$ unchanged. 
However, when a set $\Omega$ of $L \geq n+m$ true cost entries is 
available, $\varepsilon$ becomes identifiable jointly with the gauge 
parameters $f, g$, by fitting the linear relation 
$\varepsilon \, C'_{ij} + f_i + g_j = C_{ij}$ on the observed entries.

\begin{figure}
    \centering
    \includegraphics[width=1\linewidth]{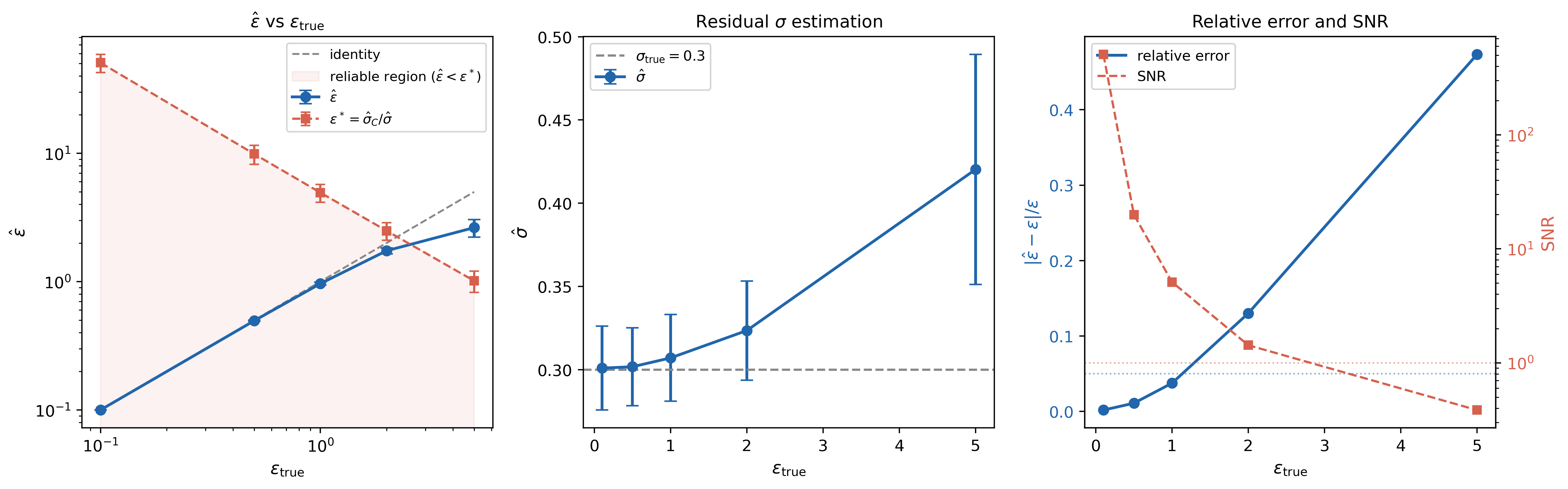}
    \caption{\textbf{Accuracy and feasibility of estimating $\epsilon$.} Left panel: estimated $\hat{\epsilon}$ vs the true $\epsilon$ (blue line). The estimation remain accurate as long as $\epsilon$ remain in the reliable zone (shaded area) defined by $\epsilon^*$ (red line). Central panel: estimation of noise level $\hat{\sigma}$ (blue line) in the transport plan $W$ as a function of the  true temperature $\epsilon$, compared to the true value (dashed line). Right panel: relative error in the estimation of the temperature (blue line) compared to the signal to noise ratio SNR (red line).}
    \label{fig:epsilon}
\end{figure}

In the noiseless setting, this estimator recovers $\varepsilon$ 
exactly for any true value and any 
$L \geq n+m$. 
When $W$ is corrupted by multiplicative noise, however, the estimator 
exhibits a systematic bias: $C'$ itself is a noisy proxy for $C/
\varepsilon$, and regressing the true cost on a noisy regressor 
produces the classical \emph{attenuation} of errors-in-variables 
regression~\cite{fuller1987measurement,carroll2006measurement}, which 
shrinks $\hat\varepsilon$ towards zero (see Appendix D). The problem arises when we have no information on the level of noise, in this case the variance $\sigma$, and $\epsilon$. Both affect the estimation of the other parameter.
The bias grows with $\varepsilon$ itself, since higher temperatures 
correspond to a lower signal-to-noise ratio in the observed plan; 
Figure~\ref{fig:epsilon} (left) shows that the estimated $\hat\varepsilon$ 
tracks the true value closely for small $\varepsilon$, but saturates 
and departs from the identity line as $\varepsilon$ increases.

Because the noise level $\sigma$ is generally unknown in practice, 
we ask whether the reliability of the estimate can be assessed from 
the data alone, without knowledge of the true $\varepsilon$.
The residuals of the fit provide exactly this information: their 
magnitude is directly related to $\sigma$ (see Appendix 
D), allowing an estimate 
$\hat\sigma$ to be extracted from the same regression used to compute 
$\hat\varepsilon$ (Figure~\ref{fig:epsilon} center). 
Combining $\hat\sigma$ with the empirical spread of $C'$ on the 
observed entries yields a data-driven reliability threshold 
$\varepsilon^* = \mathrm{std}(C'|_\Omega)/\hat\sigma = \hat{\sigma_c}/\hat\sigma $: 
the estimate is reliable when $\hat\varepsilon \ll \varepsilon^*$, 
and degrades as $\hat\varepsilon$ approaches and exceeds $\varepsilon^*$.
This threshold has a natural interpretation as a signal-to-noise 
ratio, $\mathrm{SNR} = \varepsilon^*/\hat\varepsilon$, comparing the 
scale of the true signal in $C'$ to the estimated noise floor.
Figure~\ref{fig:epsilon} (right) shows that the relative error on 
$\hat\varepsilon$ remains below a few percent as long as 
$\mathrm{SNR} \gtrsim 5$, and grows sharply once the SNR approaches 
unity --- precisely where $\hat\varepsilon$ crosses $\varepsilon^*$ 
in the left panel.
Since both $\hat\sigma$ and $\varepsilon^*$ are computable directly 
from the observed data, without any knowledge of the true $\varepsilon$ 
or $\sigma$, this provides a practical, self-contained diagnostic for 
assessing the trustworthiness of the temperature estimate in any 
given application.

Taken together, these results show that, even though the temperature 
$\varepsilon$ cannot be recovered from the transport plan alone, a 
modest amount of side information on the true cost is sufficient to 
identify it reliably, and the same regression that estimates 
$\varepsilon$ also certifies, from the data alone, the regime in 
which that estimate can be trusted --- mirroring the broader theme of 
this work, where gauge-invariant quantities can be recovered exactly, 
while the residual scale ambiguity is resolved, and its reliability 
assessed, only through additional information beyond the plan itself.

\paragraph{A general recipe for gauge-fixed cost recovery.}

The two models considered in this work --- Sinkhorn and the SubOT 
network ensemble (see Appendix C)--- are instances of the general class of 
Lagrangian-constrained transport models described by 
Eq.~\eqref{eq:general_link}. This suggests a model-agnostic recipe for 
recovering $C$ from an observed plan $W$, applicable whenever the 
generative model of $W$ admits an invertible link function $F$: 
(i) identify $F$ from the stationarity condition of the model's 
Lagrangian with respect to the plan, (ii) invert it entrywise to 
obtain $F^{-1}(W)$, and (iii) apply double centering to eliminate the 
row- and column-wise dual potentials, yielding $C$ up to the residual 
gauge freedom.

The choice of $F$ has direct consequences for the robustness of the 
resulting estimator, as illustrated by the comparison between the two 
models studied here (see Appendix C). For Sinkhorn, $F^{-1} = -\varepsilon\log(\cdot)$ 
converts multiplicative noise on $W$ into \emph{exactly} additive noise 
in the linearized space, for any noise level, since 
$\log(ab) = \log a + \log b$ holds identically. For SubOT, 
$F^{-1} = 1/(\beta \cdot)$ has no such exact linearizing property with 
respect to multiplicative perturbations of $W$: instead, noise is only 
approximately linearized, with a local amplification factor 
$1/W_{ij}^2$ that depends on the magnitude of each entry. 
The choice of link function $F$ --- determined entirely by the 
generative model one wishes to invert, not by the estimator itself 
--- sets an intrinsic bound on how favorably noise propagates through 
the recovery procedure. Models whose link function admits an exact 
algebraic identity for the noise process of interest (as is the case 
for the exponential link under multiplicative noise) will generically 
support more robust cost recovery than models lacking such an 
identity, independent of implementation details.

\section*{Conclusion and Discussion}

This work addresses a central obstacle to solving the inverse (entropic) 
optimal transport problem: the gauge freedom that renders the cost 
matrix identifiable only up to an equivalence class of additive row- 
and column-wise shifts. We showed that this ambiguity, far from being 
an obstacle to be circumvented, can be resolved exactly and in closed 
form. Double centering selects a unique, canonical 
representative of the cost equivalence class, and for many regularized OT this 
representative is recovered exactly from the observed transport plan without iteration, optimization, or any 
assumption on the structure of $C$.

Building on this exact solution, we characterized the robustness of 
cost recovery to measurement noise on the observed plan for Sinkhorn EOT. Because the 
estimator is linear in $\log W$ after double centering, multiplicative 
log-normal noise propagates into the recovered cost through an exact, 
closed-form expression, yielding an analytic prediction for the 
expected reconstruction error that matches numerical simulations 
across the full range of noise levels tested. A particularly 
consequential special case of this result is the exact invariance to 
node-wise multiplicative errors: perturbations of the form 
$\widetilde{W}_{ij} = W_{ij}\alpha_i\beta_j$, which naturally arise from 
misspecified or mismeasured marginals, leave the recovered cost 
completely unaffected. This is a genuine identifiability guarantee, not 
an approximation, and it clarifies which components of measurement 
error matter for cost recovery and which do not: only the rank-3-and-
above, non-factorizable component of the noise is visible to the 
estimator, out of the full $nm$ degrees of freedom available to a 
generic perturbation.

We further showed that partial knowledge of the true cost matrix can be 
incorporated directly into this gauge-fixed framework, by fitting the 
residual gauge freedom to a set of $L$ known entries. This recovers the 
known entries exactly whenever $L \leq n+m-1$ and the sampled entries 
span a cycle-free (tree-like) subgraph of the underlying bipartite 
structure, and degrades gracefully into a least-squares fit beyond this 
threshold. The identification of independent cycles in the bipartite 
graph of observed entries, rather than the raw count $L$, as the 
primary determinant of fitting accuracy is, to our knowledge, a novel 
structural characterization of this partial-information setting, and it 
extends naturally to the related problem of cost recovery when the 
transport plan itself is only partially observed. In both settings, the 
same graph-theoretic principle governs the trade-off between the 
information gained from additional observations and the redundancy 
that noisy or inconsistent data introduces into the gauge-fitting 
problem.

A further consequence of the gauge-fixed formulation is that it enables 
principled estimation of the entropic temperature $\varepsilon$ itself, 
which is otherwise unidentifiable from the transport plan alone. 
This closes a gap that is often left implicit in applications of 
entropic OT, where the temperature is typically treated as a fixed, 
externally chosen hyperparameter rather than as a quantity to be 
inferred alongside the cost. With our method, the temperature becomes a 'physical' parameter that can be measured in each system, enabling a deeper characterization of the optimization process.
Given 
a modest number of known cost entries, $\varepsilon$ becomes jointly 
identifiable with the gauge parameters and it is recovered exactly in absence of noise.  When the observed transposrtation plan is affected by errors, we derived a data-driven 
diagnostic -- a signal-to-noise ratio computable entirely from the 
observed plan and the known entries -- that certifies when this 
estimate can be trusted, without requiring knowledge of the true noise 
level. 

Finally, we showed that the gauge-fixing strategy developed here is not 
specific to the Sinkhorn algorithm, but extends to any transport model 
whose defining Lagrangian relates the plan to the cost through an 
invertible link function $F$, of which entropic OT and the SubOT 
network ensemble are two concrete instances. This generalization 
clarifies that the \emph{robustness} of cost recovery -- as opposed to 
its exactness, which holds for any invertible $F$ -- is governed by an 
algebraic property of $F$ itself.
This observation offers a general design principle for future 
extensions of this framework to other sub-optimal transport models: the 
choice of regularization scheme is not merely a matter of computational 
convenience or interpretability, but directly determines how favorably 
measurement noise on the observed plan propagates into the recovered 
cost.

Taken together, these results indicate that a substantial part of the 
inverse OT problem -- the recovery of the interaction structure of the 
cost, net of the unavoidable gauge and scale ambiguities of the forward 
problem -- can be solved exactly and efficiently, in closed form, 
without the iterative machinery typically employed in the literature. 
This has practical implications for applications in which the 
transport plan is observed at scale, such as trade or migration 
networks, ecological interaction data, or large urban mobility systems, 
where the computational cost of gradient-based or Bayesian inverse OT 
schemes can become prohibitive. In such settings, the method proposed 
here provides an exact, or exact-up-to-partial-information, alternative 
whose accuracy and reliability can be assessed directly from the data 
through the diagnostics developed in this work.

Several aspects of this work suggest natural directions for further 
investigation. We have 
considered marginal distributions and cost matrices without any 
additional structure (e.g., low-rank, sparse, or smooth cost 
functions); incorporating such structure, when available, could 
further constrain the residual gauge freedom beyond what is achievable 
with generic partial information on $C$, potentially reducing the 
$L \geq n+m-1$ threshold identified here. Second, the extension to the 
SubOT model considered in this work is a first step toward a broader 
class of maximum-entropy and finite-temperature transport ensembles; 
a systematic classification of which link functions $F$ admit exact 
noise-linearizing identities, and for which noise processes, would 
place the robustness comparisons presented here on a more general 
footing and could guide the choice of generative model in applications 
where measurement noise is a primary concern. Finally, while the 
present analysis focuses on synthetic data with known ground truth, the next steps will be to apply this method to real data and processes.

\section*{Author contributions}
 D.M., and R.P. carried out the numerical simulations. A.P., L.B., D.M. and R.P. developed the analytical framework. All authors contributed equally to the development of the study, writing and the revision of the manuscript.

\section*{Competing interests}
The authors declare no competing interests.
\printbibliography

@InProceedings{Chiu2022,
  title = 	 {Discrete Probabilistic Inverse Optimal Transport},
  author =       {Chiu, Wei-Ting and Wang, Pei and Shafto, Patrick},
  booktitle = 	 {Proceedings of the 39th International Conference on Machine Learning},
  pages = 	 {3925--3946},
  year = 	 {2022},
  editor = 	 {Chaudhuri, Kamalika and Jegelka, Stefanie and Song, Le and Szepesvari, Csaba and Niu, Gang and Sabato, Sivan},
  volume = 	 {162},
  series = 	 {Proceedings of Machine Learning Research},
  month = 	 {17--23 Jul},
  publisher =    {PMLR},
  url = 	 {https://proceedings.mlr.press/v162/chiu22b.html},
}

@book{Villani2009,
  title = {Optimal Transport},
  ISBN = {9783540710509},
  ISSN = {0072-7830},
  url = {http://dx.doi.org/10.1007/978-3-540-71050-9},
  DOI = {10.1007/978-3-540-71050-9},
  journal = {Grundlehren der mathematischen Wissenschaften},
  publisher = {Springer Berlin Heidelberg},
  author = {Villani,  Cédric},
  year = {2009}
}

@article{Young2021,
  title = {Reconstruction of plant–pollinator networks from observational data},
  volume = {12},
  ISSN = {2041-1723},
  url = {http://dx.doi.org/10.1038/s41467-021-24149-x},
  DOI = {10.1038/s41467-021-24149-x},
  number = {1},
  journal = {Nature Communications},
  publisher = {Springer Science and Business Media LLC},
  author = {Young,  Jean-Gabriel and Valdovinos,  Fernanda S. and Newman,  M. E. J.},
  year = {2021},
  month = jun 
}

@article{buffa2025maximum,
  title={Maximum entropy modeling of Optimal Transport: the sub-optimality regime and the transition from dense to sparse networks},
  author={Buffa, Lorenzo and Mazzilli, Dario and Piombo, Riccardo and Saracco, Fabio and Cimini, Giulio and Patelli, Aurelio},
  journal={arXiv preprint arXiv:2504.10444},
  year={2025}
}

@inproceedings{genevay2018learning,
  title={Learning generative models with sinkhorn divergences},
  author={Genevay, Aude and Peyr{\'e}, Gabriel and Cuturi, Marco},
  booktitle={International Conference on Artificial Intelligence and Statistics},
  pages={1608--1617},
  year={2018},
  organization={PMLR}
}

@article{liu2013perturbation,
  title={Perturbation of transportation polytopes},
  author={Liu, Fu},
  journal={Journal of Combinatorial Theory, Series A},
  volume={120},
  number={7},
  pages={1539--1561},
  year={2013},
  publisher={Elsevier}
}

@article{Dupuy2019,
  title = {Estimating matching affinity matrices under low-rank constraints},
  volume = {8},
  ISSN = {2049-8772},
  url = {http://dx.doi.org/10.1093/imaiai/iaz015},
  DOI = {10.1093/imaiai/iaz015},
  number = {4},
  journal = {Information and Inference: A Journal of the IMA},
  publisher = {Oxford University Press (OUP)},
  author = {Dupuy,  Arnaud and Galichon,  Alfred and Sun,  Yifei},
  year = {2019},
  month = aug,
  pages = {677–689}
}

@article{Carlier2022,
  title = {SISTA: Learning Optimal Transport Costs under Sparsity Constraints},
  volume = {76},
  ISSN = {1097-0312},
  url = {http://dx.doi.org/10.1002/cpa.22047},
  DOI = {10.1002/cpa.22047},
  number = {9},
  journal = {Communications on Pure and Applied Mathematics},
  publisher = {Wiley},
  author = {Carlier,  Guillaume and Dupuy,  Arnaud and Galichon,  Alfred and Sun,  Yifei},
  year = {2022},
  month = mar,
  pages = {1659–1677}
}

@article{Ma2020,
  doi = {10.48550/ARXIV.2002.09650},
  url = {https://arxiv.org/abs/2002.09650},
  author = {Ma,  Shaojun and Sun,  Haodong and Ye,  Xiaojing and Zha,  Hongyuan and Zhou,  Haomin},
  title = {Learning Cost Functions for Optimal Transport},
  publisher = {arXiv},
  year = {2020},
  copyright = {arXiv.org perpetual,  non-exclusive license}
}

@article{stock2021optimal,
  title={Optimal transportation theory for species interaction networks},
  author={Stock, Michiel and Poisot, Timoth{\'e}e and De Baets, Bernard},
  journal={Ecology and Evolution},
  volume={11},
  number={8},
  pages={3841--3855},
  year={2021},
  publisher={Wiley}
}

@book{fuller1987measurement,
  title={Measurement Error Models},
  author={Fuller, Wayne A.},
  year={1987},
  publisher={John Wiley \& Sons}
}

@book{carroll2006measurement,
  title={Measurement Error in Nonlinear Models: A Modern Perspective},
  author={Carroll, Raymond J. and Ruppert, David and Stefanski, Leonard A. and Crainiceanu, Ciprian M.},
  edition={2nd},
  year={2006},
  publisher={Chapman and Hall/CRC}
}

@article{rubner2000earth,
  title={The Earth Mover's Distance as a Metric for Image Retrieval},
  author={Rubner, Yossi and Tomasi, Carlo and Guibas, Leonidas J.},
  journal={International Journal of Computer Vision},
  volume={40},
  number={2},
  pages={99--121},
  year={2000},
  publisher={Springer}
}

@inproceedings{kusner2015word,
  title={From Word Embeddings To Document Distances},
  author={Kusner, Matt and Sun, Yu and Kolkin, Nicholas and Weinberger, Kilian},
  booktitle={Proceedings of the 32nd International Conference on Machine Learning},
  editor={Bach, Francis and Blei, David},
  volume={37},
  series={Proceedings of Machine Learning Research},
  pages={957--966},
  year={2015},
  publisher={PMLR}
}

@article{Li2018,
  doi = {10.48550/ARXIV.1802.03644},
  url = {https://arxiv.org/abs/1802.03644},
  author = {Li,  Ruilin and Ye,  Xiaojing and Zhou,  Haomin and Zha,  Hongyuan},
  title = {Learning to Match via Inverse Optimal Transport},
  publisher = {arXiv},
  year = {2018},
  copyright = {arXiv.org perpetual,  non-exclusive license}
}

@article{Sinkhorn1967,
  title = {Concerning nonnegative matrices and doubly stochastic matrices},
  volume = {21},
  ISSN = {0030-8730},
  url = {http://dx.doi.org/10.2140/pjm.1967.21.343},
  DOI = {10.2140/pjm.1967.21.343},
  number = {2},
  journal = {Pacific Journal of Mathematics},
  publisher = {Mathematical Sciences Publishers},
  author = {Sinkhorn,  Richard and Knopp,  Paul},
  year = {1967},
  month = may,
  pages = {343–348}
}

@inproceedings{Cuturi2013,
author = {Cuturi, Marco},
title = {Sinkhorn distances: lightspeed computation of optimal transport},
DOI = {10.5555/2999792.2999868},
year = {2013},
publisher = {Curran Associates Inc.},
address = {Red Hook, NY, USA},
booktitle = {Proceedings of the 27th International Conference on Neural Information Processing Systems - Volume 2},
pages = {2292–2300},
numpages = {9},
location = {Lake Tahoe, Nevada},
series = {NIPS'13}
}

@inproceedings{muzellec2017tsallis,
  title={Tsallis regularized optimal transport and ecological inference},
  author={Muzellec, Boris and Nock, Richard and Patrini, Giorgio and Nielsen, Frank},
  booktitle={Proceedings of the AAAI Conference on Artificial Intelligence},
  volume={31},
  number={1},
  year={2017}
}

@article{stuart2020inverse,
  title={Inverse optimal transport},
  author={Stuart, Andrew M. and Wolfram, Marie-Therese},
  journal={SIAM Journal on Applied Mathematics},
  volume={80},
  number={1},
  pages={599--619},
  year={2020},
  publisher={SIAM},
  doi={10.1137/19M1261122}
}

@article{bao2026well,
  title={Well-posedness and efficient algorithms for inverse optimal transport with bregman regularization},
  author={Bao, Chenglong and Li, Zanyu and Yang, Yunan},
  journal={Inverse Problems},
  volume={42},
  number={4},
  pages={045018},
  year={2026},
  publisher={IOP Publishing}
}

@article{baybusinov2026grand,
  title={A grand-canonical solution to a class of random optimization problems},
  author={Baybusinov, Izat B and Fenoaltea, Enrico Maria and Han, Zhen and Zhang, Yi-Cheng},
  journal={Chaos, Solitons \& Fractals},
  volume={207},
  pages={118012},
  year={2026},
  publisher={Elsevier}
}

@misc{monge1781memoire,
  title={M{\'e}moire sur la Th{\'e}orie des D{\'e}blais et des Remblais},
  author={Monge, Gaspard},
  year={1781},
  note={Histoire de l'Acad{\'e}mie Royale des Sciences de Paris, pp. 666--704}
}

@article{kantorovich1942translocation,
  title={On the translocation of masses},
  author={Kantorovich, Leonid V.},
  journal={C.R. (Doklady) Acad. Sci. URSS},
  volume={37},
  pages={199--201},
  year={1942}
}

@article{ibrahim2021optimal,
  title={Optimal transport in multilayer networks for traffic flow optimization},
  author={Ibrahim, Abdullahi Adinoyi and Lonardi, Alessandro and De Bacco, Caterina},
  journal={arXiv preprint arXiv:2106.07202},
  year={2021}
}

@article{cuturi2014ground,
  title={Ground metric learning},
  author={Cuturi, Marco and Avis, David},
  journal={Journal of Machine Learning Research},
  volume={15},
  number={1},
  pages={533--564},
  year={2014}
}

@inproceedings{kerdoncuff2021metric,
  title={Metric Learning in Optimal Transport for Domain Adaptation},
  author={Kerdoncuff, Tanguy and Emonet, R{\'e}mi and Sebban, Marc},
  booktitle={International Joint Conference on Artificial Intelligence (IJCAI)},
  year={2021}
}

@article{schiebinger2019optimal,
  title={Optimal-transport analysis of single-cell gene expression identifies developmental trajectories in reprogramming},
  author={Schiebinger, Geoffrey and Shu, Jian and Tabaka, Marcin and Cleary, Brian and Subramanian, Vidya and Solomon, Aryeh and Gould, Joshua and Liu, Siyan and Lin, Stacie and Berube, Peter and others},
  journal={Cell},
  volume={176},
  number={4},
  pages={928--943},
  year={2019},
  publisher={Elsevier}
}

@book{galichon2016optimal,
  title={Optimal Transport Methods in Economics},
  author={Galichon, Alfred},
  year={2016},
  publisher={Princeton University Press}
}

@book{bernot2008optimal,
  title={Optimal Transportation Networks: Models and Theory},
  author={Bernot, Marc and Caselles, Vicent and Morel, Jean-Michel},
  series={Lecture Notes in Mathematics},
  volume={1955},
  year={2008},
  publisher={Springer}
}

@book{peyre2019computational,
  title={Computational Optimal Transport: With Applications to Data Science},
  author={Peyr{\'e}, Gabriel and Cuturi, Marco},
  series={Foundations and Trends in Machine Learning},
  year={2019},
  publisher={Now Publishers}
}

@article{borenstein1989hubs,
  author  = {Borenstein, Severin},
  title   = {Hubs and High Fares: Dominance and Market Power in the U.S. Airline Industry},
  journal = {The RAND Journal of Economics},
  volume  = {20},
  number  = {3},
  pages   = {344--365},
  year    = {1989},
  doi     = {10.2307/2555575}
}

\subsection*{Appendix A: Robustness to noise}
\label{appendix:noise}

In practice, $W$ is never observed exactly: empirical transport plans 
are subject to measurement noise, finite-sample fluctuations, or 
model misspecification. 
We therefore study the robustness of the estimator~\eqref{eq:estimator} 
by analyzing how errors in the observed plan $W$ propagate into errors 
in the recovered cost $C'$.
To this end, we consider a multiplicative noise model
\begin{equation}
    \widetilde{W}_{ij} = W_{ij} \cdot \eta_{ij},
    \label{eq:noise_model}
\end{equation}
where $\eta_{ij}$ are i.i.d.\ log-normal random variables, 
$\log \eta_{ij} \sim \mathcal{N}(0, \sigma^2)$, and $\sigma > 0$ 
controls the noise level.
Multiplicative noise is a natural choice here: it preserves the 
non-negativity of the transport plan, it acts proportionally to the 
magnitude of each entry, and it is the canonical noise model for 
quantities defined in log-space, as is the case for entropic transport 
plans.
We do not re-project $\widetilde{W}$ onto the transportation polytope 
with fixed marginals $s$ and $\sigma$, as we assume the true marginals 
are not available: in the observational setting we consider, both the 
noisy plan and its marginals are inferred directly from the data.

\subsubsection*{Exact propagation of noise}

The key observation is that the estimator~\eqref{eq:estimator} is 
linear in $\log W$ after double centering. Applying it to the noisy 
plan $\widetilde{W}$ gives
\begin{equation}
    \widetilde{C}' 
    = -(\log \widetilde{W})' 
    = -(\log W + \log H)',
    \label{eq:noisy_estimator}
\end{equation}
where $H_{ij} = \log \eta_{ij} \sim \mathcal{N}(0,\sigma^2)$ is the 
matrix of log-noise terms.
Since double centering is a linear operator and $C' = -(\log W)'$, 
the reconstruction error is exactly
\begin{equation}
    \widetilde{C}' - C' = -H',
    \label{eq:error}
\end{equation}
where $H' $ is the double-centered log-noise matrix.
This is a remarkable simplification: the error depends \emph{only} on 
the noise realization $H$, and not on the true cost $C$ or the true 
plan $W$. 

\subsubsection*{Analytical characterization of the error}

Since double centering is an orthogonal projection onto the subspace 
of matrices with zero row and column means, which has dimension 
$(n-1)(m-1)$, and since the entries of $H$ are i.i.d.\ 
$\mathcal{N}(0, \sigma^2)$, the expected squared Frobenius norm of 
the error is
\begin{equation}
    \mathbb{E}\left[\| \widetilde{C}' - C' \|_F^2\right] 
    = \sigma^2 (n-1)(m-1).
    \label{eq:expected_error}
\end{equation}
To measure the reconstruction quality independently of the scale of 
$C$, we define the relative error
\begin{equation}
    d_{\mathrm{rel}} 
    = \frac{\|\widetilde{C}' - C'\|_F}{\|C'\|_F},
    \label{eq:rel_error}
\end{equation}
whose expected value is
\begin{equation}
    \mathbb{E}[d_{\mathrm{rel}}] 
    \approx \frac{\sigma \sqrt{(n-1)(m-1)}}{\|C'\|_F}.
    \label{eq:expected_rel_error}
\end{equation}
This expression has a transparent interpretation: the relative 
reconstruction error grows linearly with the noise level $\sigma$, 
scales with the size of the problem through the factor 
$\sqrt{(n-1)(m-1)}$, and is inversely proportional to the 
\emph{signal strength} $\|C'\|_F$, i.e.\ the magnitude of the 
true cost in the gauge-fixed representation.

To quantify the noise injected into the transport plan independently 
of the matrix size, we use the log-space distance
\begin{equation}
    d_{\log}(W, \widetilde{W}) 
    = \|H\|_F 
    = \|\log W - \log \widetilde{W}\|_F,
    \label{eq:noise_distance}
\end{equation}
which satisfies $\mathbb{E}[d_{\log}^2] = \sigma^2 nm$ and is the 
natural metric for multiplicative perturbations.
Combining~\eqref{eq:expected_error} and~\eqref{eq:noise_distance}, 
the expected reconstruction error can be written as
\begin{equation}
    \mathbb{E}\left[\|\widetilde{C}' - C'\|_F^2\right] 
    = \frac{(n-1)(m-1)}{nm} \cdot 
      \mathbb{E}\left[d_{\log}(W,\widetilde{W})^2\right],
    \label{eq:error_vs_noise}
\end{equation}
showing that only the fraction $(n-1)(m-1)/nm$ of the injected noise 
is visible to the estimator, while the remaining fraction, 
corresponding to the gauge directions, is automatically discarded 
by double centering.

Taken together, these two results delineate a sharp boundary in the 
space of measurement errors affecting $W$: perturbations that are 
exactly factorized across sources and targets leave the recovered 
cost completely unaffected, while any non-factorized component of 
the noise propagates into the reconstruction with a magnitude that 
is fully predictable from the noise level alone, independently of 
the underlying true cost matrix.

\begin{figure}[htb]
    \centering
    \includegraphics[width=1\linewidth]{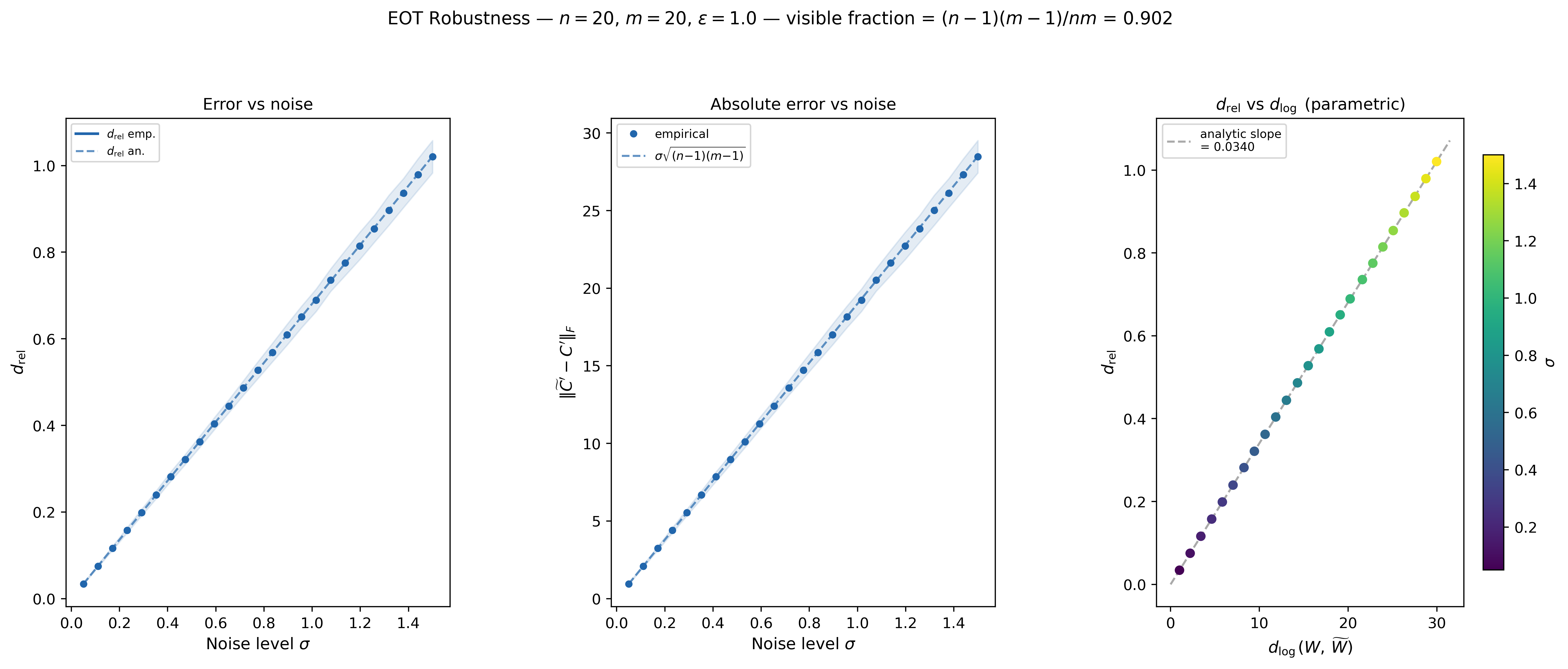}
    \caption{\bf{Numerical validation of noise}}
    \label{fig:robustness}
\end{figure}

We validate the analytical predictions above with numerical 
experiments. For a range of noise levels $\sigma$, we generate 
noisy plans $\widetilde{W}$ according to~\eqref{eq:noise_model}, 
apply the estimator~\eqref{eq:estimator}, and measure both 
$d_{\log}(W, \widetilde{W})$ and $d_{\mathrm{rel}}$.
Figure~\ref{fig:robustness} shows $d_{\mathrm{rel}}$ and 
$d_{\log}$ as a function of $\sigma$, averaged over multiple 
noise realizations. Both quantities grow linearly with $\sigma$, 
in agreement with~\eqref{eq:expected_rel_error} 
and~\eqref{eq:noise_distance}, and the analytical curves provide 
an accurate description of the empirical results across the full 
range of noise levels considered.

A particularly important class of measurement errors are 
\emph{node-wise} multiplicative perturbations, in which each entry 
of the observed plan is corrupted by source- and target-specific 
factors:
\begin{equation}
    \widetilde{W}_{ij} = W_{ij} \cdot \alpha_i \cdot \beta_j,
    \label{eq:nodewise}
\end{equation}
for arbitrary strictly positive vectors $(\alpha_i)$ and $(\beta_j)$.
Such errors arise naturally in practice whenever the marginal 
distributions $s$ and $\sigma$ are themselves measured with 
multiplicative noise, or whenever the observation process introduces 
systematic row- and column-specific biases — for instance, 
heterogeneous detection efficiencies across sources and targets.
In log-space, perturbation~\eqref{eq:nodewise} takes the additive form
$\log \widetilde{W}_{ij} = \log W_{ij} + \log \alpha_i + \log \beta_j$,
which lies entirely in the gauge subspace spanned by source- and 
target-specific additive terms.
Since the double-centering operator projects orthogonally onto the 
complement of this subspace, it annihilates the perturbation exactly, 
and the estimator satisfies
\begin{equation}
    \widetilde{C}' = -(\log \widetilde{W})' = -(\log W)' = C'
\end{equation}
identically, regardless of the magnitude of $\alpha_i$ and $\beta_j$.
The recovery of $C'$ is therefore \emph{exactly} robust to node-wise 
multiplicative errors, not merely approximately so.
This result can be seen as a strong identifiability guarantee: 
even if the marginals $s$ and $\sigma$ are completely misspecified, 
the interaction structure of the cost matrix remains perfectly 
recoverable from the observed plan.

More generally, this invariance result has implications for the 
effective rank of any perturbation that is visible to the estimator.
Consider a generic multiplicative noise matrix in log-space, 
$H = \log \widetilde{W} - \log W$, of rank $r$. 
The double-centering operator projects $H$ onto the orthogonal 
complement of the gauge subspace, which is spanned by all matrices 
of the form $f_i + g_j$.
This gauge subspace has dimension $n + m - 1$ and contains, in 
particular, all rank-1 matrices of the form $u \mathbf{1}^T$ and 
$\mathbf{1} v^T$, as well as their linear combinations, which form 
a subspace of rank at most 2.
The projection therefore removes at least a rank-2 component from 
$H$, so that the perturbation visible to the estimator, $H' = 
\mathrm{dc}(H)$, has rank at most $r - 2$ in general, and exactly 
zero whenever $r \leq 2$.

This has a direct and practically relevant consequence: any 
measurement error that can be expressed as a rank-2 multiplicative 
perturbation in log-space — including all node-wise errors 
\eqref{eq:nodewise} as a special case — is \emph{completely 
invisible} to the estimator.
Conversely, the minimal perturbation rank that can affect the 
recovered cost is 3, and even then only the component orthogonal 
to the gauge subspace contributes to the reconstruction error.
This provides a precise characterization of the \emph{effective 
degrees of freedom} of the noise with respect to the estimator: 
out of the $nm$ degrees of freedom of a generic perturbation matrix, 
only $(n-1)(m-1)$ are visible, corresponding to the dimension of 
the double-centered subspace, and the remaining $n + m - 1$ are 
absorbed by the gauge invariance regardless of their magnitude.

\subsection*{Appendix B: Missing values}
\paragraph{Partial double-centering with missing entries.}
When a subset of entries of $W$ is unobserved, we cannot apply the 
standard double-centering operator, which requires row and column 
means computed over the full matrix. We instead define a partial 
double-centering that uses only the retained entries: given a mask 
$\Omega \subseteq \{1,\dots,n\} \times \{1,\dots,m\}$ of observed 
positions, we set
\begin{equation}
    C'_{ij} = -\log W_{ij} 
    - \overline{(-\log W)}_{i\cdot}^{\Omega} 
    - \overline{(-\log W)}_{\cdot j}^{\Omega} 
    + \overline{(-\log W)}_{\cdot\cdot}^{\Omega}
    \quad \forall (i,j) \in \Omega,
    \label{eq:partial_dc}
\end{equation}
where $\overline{(-\log W)}_{i\cdot}^{\Omega}$ and 
$\overline{(-\log W)}_{\cdot j}^{\Omega}$ denote the row and column 
means of $-\log W$ computed only over the observed entries in row 
$i$ and column $j$ respectively, and 
$\overline{(-\log W)}_{\cdot\cdot}^{\Omega}$ is the corresponding 
grand mean over $\Omega$. This estimator is defined only on $\Omega$; 
no value is assigned to unobserved entries.

\paragraph{Missing-data model and evaluation.}
For each trial, we generate a random cost matrix $C$ with 
$n = m = 20$ as in the previous experiments, draw marginals $s, 
\sigma$ from the uniform distribution on the simplex, compute the 
exact entropic plan $W$ via Sinkhorn, and apply multiplicative 
log-normal noise with $\sigma_{\text{noise}} = 0.3$ to obtain 
$\widetilde{W}$. We then remove $M$ entries from $\widetilde{W}$ 
uniformly at random without replacement, and apply the partial 
double-centering estimator~\eqref{eq:partial_dc} to the remaining 
$nm - M$ entries. We vary $M$ from $0$ to $nm/3$.
Results are averaged over $100$ independent trials for each value of 
$M$, with $C$, the noise realization, and the missing-entry mask 
resampled at each trial.

\paragraph{Marginal error.}
To relate the reconstruction error to the distortion induced in the 
implicitly estimated marginals, at each trial we compute the 
empirical marginals from the retained entries of $\widetilde{W}$, 
\begin{equation}
    \hat\mu_i = \sum_{j \,:\, (i,j)\in\Omega} \widetilde{W}_{ij}, 
    \qquad
    \hat\nu_j = \sum_{i \,:\, (i,j)\in\Omega} \widetilde{W}_{ij},
\end{equation}
and report their relative error with respect to the true marginals, 
$\|\hat\mu - s\|_2 / \|s\|_2$ (and analogously for $\hat\nu$ and 
$\sigma$), averaged over rows and columns.

\paragraph{Cycle count.}
As in the partial-information experiments, we characterize the 
structure of the retained entries via the bipartite graph 
$G_\Omega = (V_r \cup V_c, \Omega)$ induced by the observed positions, 
and compute the number of independent cycles 
$\gamma(\Omega) = |\Omega| - |V_\Omega| + \kappa(\Omega)$, where 
$|V_\Omega|$ is the number of distinct rows and columns touched by 
$\Omega$ and $\kappa(\Omega)$ is the number of connected components 
of $G_\Omega$. Since $M$ missing entries are removed uniformly at 
random from the full $n\times m$ grid, $|\Omega| = nm - M$ decreases 
linearly with $M$, while $\gamma(\Omega)$ decreases correspondingly as 
the retained graph loses redundant connections.

\begin{figure}
    \centering
    \includegraphics[width = 1\linewidth]{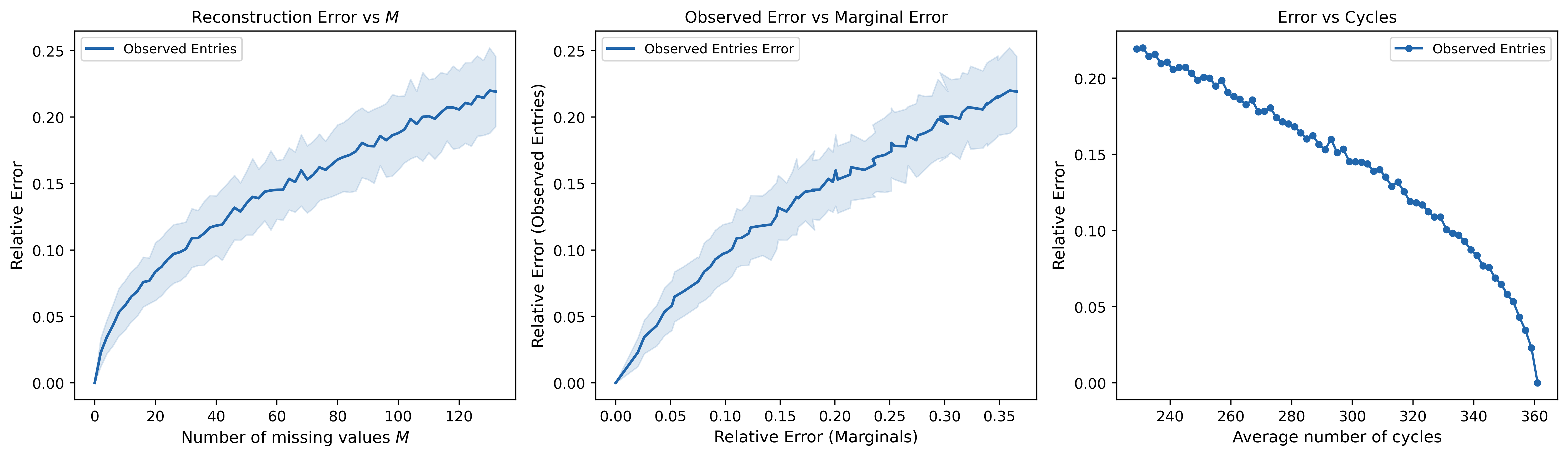}
    \caption{Error due to missing values}
    \label{fig:missing}
\end{figure}

Figure~\ref{fig:missing} (left) shows that the reconstruction error on 
the retained entries grows monotonically with the number of missing 
values $M$, but sublinearly: the error increases sharply for small 
$M$ and flattens out as $M$ grows, reflecting the fact that the row 
and column means entering the partial double-centering become 
increasingly unreliable estimates of their true (full-matrix) 
counterparts as fewer observations remain to compute them.

This connection is made precise in Figure~\ref{fig:missing} (center), 
which compares the reconstruction error to the relative error in the 
empirical marginals estimated from the incomplete plan, 
$\hat\mu_i = \sum_{j : (i,j) \text{ observed}} W_{ij}$ and 
analogously for $\hat\nu_j$, relative to the true marginals $s, 
\sigma$. The two quantities track each other closely and nearly 
linearly, showing that the error introduced by missing data in the 
cost recovery is directly inherited from the distortion that missing 
entries induce in the implicitly estimated marginals: rows and 
columns with fewer observed entries yield less reliable estimates of 
their means, which propagates directly into the double-centered cost.

Consistent with the role of the bipartite graph structure identified 
in the partial-information setting, Figure~\ref{fig:missing} (right) 
shows that the reconstruction error is well predicted by the number 
of independent cycles in the bipartite graph of retained entries: as 
$M$ decreases and the retained graph becomes denser and more 
redundant (more cycles), the error decreases towards zero, while 
sparser retained graphs (fewer cycles, occurring at larger $M$) yield 
systematically larger errors. This mirrors the mechanism identified 
earlier for partial knowledge of $C$, and confirms that graph 
connectivity and redundancy, rather than the raw count of missing 
entries alone, is the primary structural driver of reconstruction 
accuracy under missing data.

\subsection*{Appendix C: generalization to network-ensemble models of sub-optimal transport}

The estimator derived so far is specific to the entropic regularization 
of OT via the Sinkhorn algorithm, where the optimal plan takes the 
exponential form $W_{ij} = e^{f_i/\varepsilon} e^{-C_{ij}/\varepsilon} 
e^{g_j/\varepsilon}$. A natural question is whether the same 
gauge-invariant recovery strategy extends to other probabilistic 
formulations of sub-optimal transport. We consider the maximum-entropy 
network ensemble recently proposed by Buffa et al.~\cite{buffa2025maximum}, 
in which the expected weight of each edge in a bipartite network is 
given by
\begin{equation}
    \langle w_{i\alpha} \rangle = \frac{1}{\beta C_{i\alpha} + t_i + \theta_\alpha},
    \label{eq:subot_mean}
\end{equation}
where $\beta$ plays the role of an inverse temperature and $t_i, 
\theta_\alpha$ are Lagrange multipliers enforcing prescribed node 
strengths, analogous to the dual potentials of entropic OT. Unlike the 
Sinkhorn plan, which depends on $C$ through an exponential, this model 
depends on $C$ through a reciprocal (harmonic) relation.

This structural difference suggests a direct generalization of our 
recovery method: taking the reciprocal of the observed plan, rather 
than its logarithm, linearizes Eq.~\eqref{eq:subot_mean} exactly:
\begin{equation}
    \frac{1}{\langle w_{i\alpha}\rangle} = \beta C_{i\alpha} + t_i + \theta_\alpha,
    \label{eq:subot_linear}
\end{equation}
which has precisely the same additive gauge structure as the entropic 
case. Applying double centering to $1/W$ therefore recovers $\beta C$ 
exactly, up to gauge, mirroring our central result for entropic OT: 
$C'_\beta := (1/W)' = \beta\, C'$, exactly and without approximation 
in the noiseless case, for any $\beta$ and any matrix size.

\subsubsection*{Robustness comparison between models}

We compared the robustness of cost recovery between the two models 
under identical conditions: the same true cost matrix $C$, the same 
marginals, and the same injected noise on the observed plan. Because 
the two models relate $C$ to $W$ through fundamentally different 
functions, a naive comparison at matched marginals can be misleading: 
calibrating the SubOT temperature $\beta$ to match a target mean edge 
weight can inadvertently drive the model toward the entropy-dominated 
(BiWCM) regime described in ref.~\cite{buffa2025maximum}, in which the 
cost term becomes negligible and the recoverable signal $\|C'_\beta\|_F$ 
collapses to near zero --- an effect entirely decoupled from the 
robustness of the estimator itself. We therefore calibrate $\beta$ to 
match the recoverable signal strength between the two models, 
$\|C'_\beta\|_F = \|C'\|_F$, ensuring that both models encode a 
comparable amount of information about $C$ before assessing how well 
that information survives measurement noise.

\begin{figure}[htb]
    \centering
    \includegraphics[width=1\linewidth]{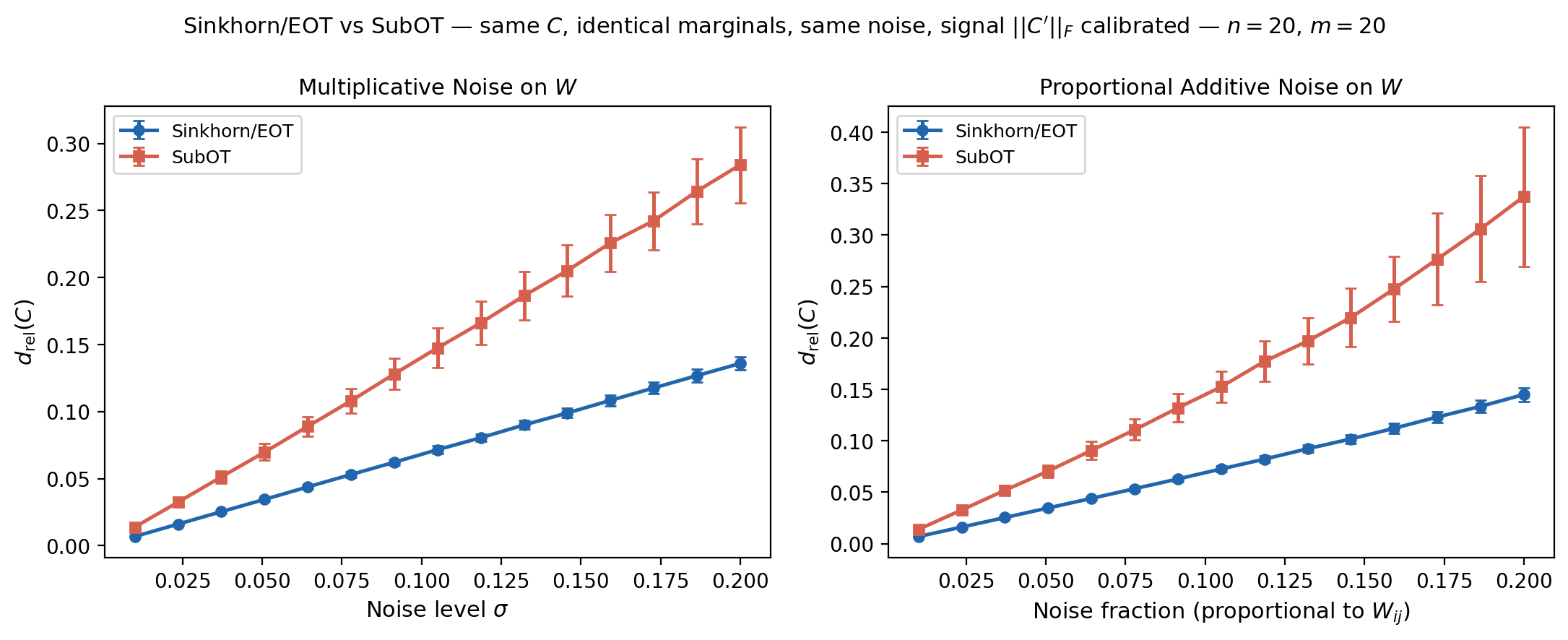}
    \caption{Error comparison between Sinkhorn and SubOT}
    \label{fig:subot_comparison}
\end{figure}

Under this matched-signal calibration, Figure~\ref{fig:subot_comparison} 
shows that the reciprocal-based estimator for the SubOT model is 
consistently, but only moderately, less robust than the entropic 
estimator: the relative reconstruction error is approximately a factor 
of two larger across the full range of noise levels tested, for both 
multiplicative and proportional additive noise on $W$. This is 
consistent with the absence, for the reciprocal transform, of an exact 
linearizing identity analogous to $\log(ab) = \log a + \log b$, which 
underlies the exact error propagation derived for the Sinkhorn: for SubOT, even small perturbations of 
$W$ are amplified locally by a factor $1/W_{i\alpha}^2$, an effect that 
is absent for the logarithmic transform. This amplification becomes 
severe when individual edge weights are close to zero, which we 
verified is primarily a consequence of poor parameter calibration 
(vanishing signal, as discussed above) rather than an intrinsic 
limitation of the model at realistic parameter values: once $\beta$ is 
calibrated to preserve a meaningful cost signal, and edge weights are 
kept away from the numerical instability threshold, the two models 
exhibit comparable and well-behaved robustness, with SubOT paying only 
a modest, consistent penalty relative to the entropic case.

\subsection*{Appendix D: numerical simulations}
\subsubsection*{Cost recovery under partial information: experimental setup}

To assess the benefit of incorporating partial knowledge of the true 
cost matrix into the gauge-fixed reconstruction, we design a numerical 
experiment in which $L$ entries of $C$ are assumed known and used to 
select a specific representative within the gauge class of $C'$.

\paragraph{Data generation.}
For each trial, we generate a random $n \times m$ cost matrix $C$ 
with $n = m = 15$, whose entries are drawn as 
$C_{ij} = A_{ij}^2 + \frac{1}{2} B_{ij}^2$, where 
$A_{ij}, B_{ij} \overset{\text{iid}}{\sim} \mathcal{N}(0,1)$, 
ensuring a strictly positive, non-trivial cost structure.
Source and target marginals $s$ and $\sigma$ are drawn independently 
from a uniform distribution on the simplex, obtained by normalizing 
$n$ and $m$ i.i.d.\ uniform random variables respectively.
The entropic optimal transport plan $W$ is computed via the Sinkhorn 
algorithm with temperature $\varepsilon = 1.0$, and multiplicative 
log-normal noise with parameter $\sigma_{\text{noise}} = 0.3$ is 
applied to $W$ to simulate measurement error, yielding the observed 
plan $\widetilde{W}$.
The gauge-fixed cost estimate is then $C' = -(\log \widetilde{W})'$, 
which differs from $C$ by both the gauge ambiguity and the 
reconstruction error due to noise.

\paragraph{Gauge fitting.}
Given $L$ observed entries $\{C_{ij}\}_{(i,j)\in\Omega}$ of the true 
cost matrix, we seek gauge parameters $f \in \mathbb{R}^n$ and 
$g \in \mathbb{R}^m$ such that
\begin{equation}
    C^*_{ij} = C'_{ij} + f_i + g_j \approx C_{ij}
    \quad \forall (i,j) \in \Omega.
    \label{eq:gauge_fit}
\end{equation}
This amounts to solving the linear system
\begin{equation}
    f_i + g_j = C_{ij} - C'_{ij} \quad \forall (i,j) \in \Omega,
    \label{eq:linear_system}
\end{equation}
which has $n+m$ unknowns but only $n+m-1 = 29$ effective degrees of 
freedom due to the residual gauge redundancy $f \to f + c$, $g \to 
g - c$.
System~\eqref{eq:linear_system} is solved via singular value 
decomposition through the Moore--Penrose pseudoinverse of the 
design matrix $A \in \mathbb{R}^{L \times (n+m)}$, where each 
row $k$ has exactly two nonzero entries: a $1$ in column $i_k$ 
and a $1$ in column $n + j_k$, corresponding to the observed 
pair $(i_k, j_k) \in \Omega$.
This yields the minimum-norm solution when the system is 
underdetermined ($L < n+m-1$) and the least-squares solution 
when it is overdetermined ($L > n+m-1$).

\paragraph{Sampling strategies.}
The structure of the bipartite graph $G_\Omega = (V_r \cup V_c, \Omega)$, 
whose nodes are the $n$ source indices $V_r$ and the $m$ target indices 
$V_c$ and whose edges are the observed pairs $(i,j) \in \Omega$, 
determines the solvability of~\eqref{eq:linear_system}.
Each independent cycle in $G_\Omega$ imposes an additional linear 
constraint on $C - C'$ that is generically not satisfied when $C'$ is 
estimated from a noisy plan, introducing a non-zero residual on the 
known entries even for $L \leq n+m-1$.
The number of independent cycles is given by
\begin{equation}
    \gamma(\Omega) = L - |V_\Omega| + \kappa(\Omega),
\end{equation}
where $|V_\Omega|$ is the number of distinct row and column indices 
appearing in $\Omega$ and $\kappa(\Omega)$ is the number of connected 
components of $G_\Omega$.
We compare two sampling strategies:
\begin{itemize}
    \item \emph{Random spanning tree}: entries are selected by first 
    constructing a random spanning tree of the complete bipartite graph 
    $K_{n,m}$, then adding $\max(0, L-(n+m-1))$ extra edges chosen 
    uniformly at random among the remaining pairs. 
    This guarantees $\gamma(\Omega) = 0$ for all $L \leq n+m-1 = 29$ 
    and $\gamma(\Omega) = L - (n+m-1)$ for $L > n+m-1$.
    \item \emph{Fully random}: $L$ entries are chosen uniformly at 
    random without replacement, with no structural constraint. 
    This introduces cycles already for small $L$, with the expected 
    number of cycles growing approximately as $L - |V_\Omega| + 
    \kappa(\Omega)$ under uniform sampling.
\end{itemize}
In both cases $L$ is varied from $1$ to $200$, covering the 
underdetermined regime ($L < 29$), the critical threshold ($L = 29$), 
and the overdetermined regime ($L > 29$) up to more than six times 
the threshold.

\paragraph{Evaluation metrics.}
For each trial we record two quantities.
The \emph{known-entry error}
\begin{equation}
    e_\Omega = \frac{\sum_{(i,j)\in\Omega} 
    (C^*_{ij} - C_{ij})^2}{\|C'\|_F^2}
\end{equation}
measures how well $C^*$ reproduces the known entries, and equals zero 
when system~\eqref{eq:linear_system} is exactly solvable.
The \emph{unknown-entry error}
\begin{equation}
    e_{\bar\Omega} = \frac{\sum_{(i,j)\notin\Omega} 
    (C^*_{ij} - C_{ij})^2}{\|C'\|_F^2}
\end{equation}
measures the quality of the reconstruction on the unobserved entries, 
which depends on how well the fitted parameters $f, g$ generalize 
beyond the observed set.
Both errors are normalized by $\|C'\|_F^2$ to make them directly 
comparable across trials with different cost scales.
Results are averaged over $N_{\text{rep}} = 300$ independent trials 
for each value of $L$, with both $C$ and $\Omega$ resampled at each 
trial.

\subsubsection*{Estimating the temperature $\varepsilon$: experimental setup}

\paragraph{Joint estimation of $\varepsilon$ and the gauge.}
Given a set $\Omega$ of $L$ known entries of the true cost matrix, we 
jointly estimate the temperature $\varepsilon$ and the gauge 
parameters $f \in \mathbb{R}^n$, $g \in \mathbb{R}^m$ by solving the 
linear system
\begin{equation}
    \varepsilon \, C'_{ij} + f_i + g_j = C_{ij} 
    \quad \forall (i,j) \in \Omega,
    \label{eq:eps_system}
\end{equation}
where $C' = -(\log W)'$ is computed once from the full observed plan 
$W$ (i.e., using all $nm$ entries, not only those in $\Omega$), so 
that $C'$ is always available as a complete, well-defined regressor. 
System~\eqref{eq:eps_system} has $1+n+m$ unknowns 
($\varepsilon, f, g$) but only $n+m$ effective degrees of freedom, 
due to the same gauge redundancy $f \to f+c$, $g \to g-c$ discussed 
in the context of partial information recovery; it is therefore 
identifiable for $L \geq n+m$.
We solve it via the Moore--Penrose pseudoinverse of the design matrix 
$A \in \mathbb{R}^{L \times (1+n+m)}$, whose first column holds the 
values $C'_{ij}$ for $(i,j) \in \Omega$ and whose remaining columns 
are row and column indicators as in the partial-information setting; 
this yields the least-squares estimate $\hat\varepsilon_{\text{OLS}}$ 
together with the residuals $r_{ij} = C_{ij} - (\hat\varepsilon_{\text{OLS}} 
C'_{ij} + \hat f_i + \hat g_j)$.
In the noiseless case ($W$ exact), this procedure recovers 
$\varepsilon$ to machine precision for any $L \geq n+m$, since 
$C'_{ij} = C_{ij}/\varepsilon$ exactly up to gauge.

\paragraph{Bias under noisy observation of $W$.}
When $W$ is corrupted by multiplicative log-normal noise with 
parameter $\sigma$ (as in the robustness experiments above), $C'$ 
itself carries a noise component: $C'_{ij} = C_{ij}/\varepsilon - 
H'_{ij}$, where $H'$ is the double-centered log-noise matrix. Since 
$C'$ appears as the regressor in~\eqref{eq:eps_system}, this is an 
errors-in-variables regression problem, and $\hat\varepsilon_{\text{OLS}}$ 
is attenuated towards zero, with the expected bias following
\begin{equation}
    \mathbb{E}[\hat\varepsilon_{\text{OLS}}] 
    = \varepsilon \cdot 
    \frac{\mathrm{Var}(C'|_\Omega)}{\mathrm{Var}(C'|_\Omega) 
    + \varepsilon^2 \sigma^2},
    \label{eq:eps_bias}
\end{equation}
which grows more severe as $\varepsilon$ increases, since higher 
temperatures correspond to a smaller signal-to-noise ratio in $W$ 
itself. We verified that attempting to invert~\eqref{eq:eps_bias} 
analytically to correct for this bias --- either via a fixed-point 
iteration or by solving the resulting quadratic equation in 
$\varepsilon$ --- does not yield a stable or accurate correction in 
practice, as the correction itself depends sensitively on 
$\hat\varepsilon_{\text{OLS}}$ and on an estimate of $\sigma$ that is 
only reliable in the same low-bias regime where correction is least 
needed. We therefore report the uncorrected estimator 
$\hat\varepsilon_{\text{OLS}}$ throughout, and instead focus on 
quantifying its reliability directly from the data.

\paragraph{Estimating the reliability of $\hat\varepsilon$.}
Under model~\eqref{eq:eps_bias}, the residuals of the fit 
in~\eqref{eq:eps_system} satisfy $r_{ij} \approx 
\hat\varepsilon_{\text{OLS}} \, H'_{ij}$, so that the noise level can 
be estimated directly from the regression as
\begin{equation}
    \hat\sigma = \frac{1}{|\hat\varepsilon_{\text{OLS}}|} 
    \sqrt{\frac{\sum_{(i,j)\in\Omega} r_{ij}^2}{L - (n+m+1)}},
    \label{eq:sigma_hat}
\end{equation}
where the denominator accounts for the degrees of freedom used in 
fitting $\varepsilon, f, g$.
Combining $\hat\sigma$ with the empirical standard deviation of $C'$ 
on the observed entries, $\mathrm{std}(C'|_\Omega)$, we define the 
data-driven reliability threshold
\begin{equation}
    \varepsilon^* = \frac{\mathrm{std}(C'|_\Omega)}{\hat\sigma},
    \label{eq:eps_star}
\end{equation}
and the associated signal-to-noise ratio 
$\mathrm{SNR} = \varepsilon^* / \hat\varepsilon_{\text{OLS}}$. 
Both $\hat\sigma$ and $\varepsilon^*$ are computable entirely from 
the observed plan $W$ and the known entries in $\Omega$, without any 
reference to the true $\varepsilon$, making this a practical 
diagnostic that can be evaluated in any real application.

\paragraph*{Experimental setup.}
We generate random cost matrices $C$ with $n=m=20$ as in the 
robustness experiments, with source and target marginals drawn from 
the uniform distribution on the simplex. For each true temperature 
$\varepsilon \in \{0.1, 0.5, 1.0, 2.0, 5.0\}$, we compute the exact 
entropic plan $W$ via Sinkhorn, apply multiplicative log-normal noise 
with $\sigma = 0.3$, and reveal $L = 3(n+m) = 120$ true entries of 
$C$ sampled via the spanning-tree procedure described above (which 
guarantees a well-conditioned, cycle-free design for $L \leq n+m-1$ 
and a connected graph with controlled redundancy beyond that 
threshold). We solve~\eqref{eq:eps_system} to obtain 
$\hat\varepsilon_{\text{OLS}}$, estimate $\hat\sigma$ 
via~\eqref{eq:sigma_hat}, and compute $\varepsilon^*$ 
via~\eqref{eq:eps_star}. Results are averaged over $300$ independent 
trials for each value of $\varepsilon$, with both $C$ and the 
sampled set $\Omega$ resampled at each trial.

\subsubsection*{SubOT model: numerical solution and calibration}

\paragraph{Fixed-point solution for the dual potentials.}
Given a cost matrix $C$, target strengths $s^*, r^*$, and a 
temperature $\beta$, the Lagrange multipliers $t, \theta$ in 
Eq.~\eqref{eq:subot_mean} are determined by the strength constraints 
(Eqs.~33--34 of ref.~\cite{buffa2025maximum}):
\begin{equation}
    s_i^* = \sum_\alpha \frac{1}{\beta C_{i\alpha} + t_i + \theta_\alpha},
    \qquad
    r_\alpha^* = \sum_i \frac{1}{\beta C_{i\alpha} + t_i + \theta_\alpha}.
\end{equation}
We solve this system via a Sinkhorn-style multiplicative fixed-point 
iteration, updating $t_i \leftarrow t_i \cdot (\hat s_i / s_i^*)$ and 
$\theta_\alpha \leftarrow \theta_\alpha \cdot (\hat r_\alpha / 
r_\alpha^*)$ in alternation, where $\hat s_i, \hat r_\alpha$ 
are the strengths implied by the current multipliers, initialized as 
$t_i = 2/s_i^*$, $\theta_\alpha = 2/r_\alpha^*$ following the 
initialization scheme of the original paper. Convergence is assessed 
via the relative error on the implied strengths, with a tolerance of 
$10^{-11}$.

\paragraph{Calibration for cross-model comparison.}
To compare robustness against the entropic estimator under matched 
conditions, we generate a single cost matrix $C$ and a single pair of 
marginals $s^*, r^*$ (drawn as log-normal node strengths with 
mean corresponding to realistic edge-count scales, i.e., individual 
weights ranging from order 1 to a few hundred), shared identically by 
both models. The entropic plan is computed via standard Sinkhorn 
iteration with temperature $\varepsilon$ on these same marginals. For 
the SubOT model, rather than calibrating $\beta$ to match a target 
mean edge weight --- which we found can inadvertently collapse the 
cost signal $\|C'_\beta\|_F$ toward zero by driving the model into the 
entropy-dominated regime --- we calibrate $\beta$ via bisection to 
match the recoverable signal strength of the entropic model, 
$\|(1/W_\beta)'\|_F = \|(-\log W_\varepsilon)'\|_F$, ensuring both 
models encode an equal amount of cost information prior to the 
introduction of noise.

\paragraph{Noise injection and evaluation.}
We inject two types of noise directly on the observed plan $W$, 
identically for both models: (i) multiplicative log-normal noise, 
$\widetilde{W}_{ij} = W_{ij}\, e^{H_{ij}}$ with $H_{ij} \sim 
\mathcal{N}(0,\sigma^2)$ and 
(ii) proportional additive Gaussian noise, $\widetilde{W}_{ij} = 
\max(W_{ij} + \delta_{ij},\, \epsilon_{\text{floor}})$ with 
$\delta_{ij} \sim \mathcal{N}(0, (\text{frac} \cdot W_{ij})^2)$, where 
the noise standard deviation scales with each entry's own magnitude 
to avoid numerical pathologies in matrices with wide dynamic range. 
For the proportional additive case, we restrict the tested noise 
fraction to a range below the threshold at which individual entries 
are driven close to zero with non-negligible probability across the 
$n \times m$ matrix and repeated trials, since beyond this threshold 
the reciprocal transform of the SubOT estimator produces occasional 
numerically catastrophic outliers that dominate the average 
reconstruction error and obscure the underlying trend. For each noise 
level, we recover $C'$ from the noisy plan using the corresponding 
model-specific estimator ($-\log(\cdot)$ for Sinkhorn, $1/(\cdot)$ 
for SubOT, both followed by double centering) and report the relative 
Frobenius error against the noiseless reference $C'$, averaged over 
$300$ independent noise realizations, with $n=m=20$.

\end{document}